# Hybrid magnon-phonon cavity for large-amplitude terahertz spin-wave excitation

Shihao Zhuang,[1] Xufeng Zhang[2], Yujie Zhu[1], Nian X. Sun,[2] Chang-Beom Eom,[1] Paul G. Evans,[1] Jia-Mian Hu[1*]

[1]Department of Materials Science and Engineering, University of Wisconsin-Madison, Madison, WI, 53706, USA

[2]Department of Electrical and Computer Engineering, Northeastern University, Boston, Massachusetts 02115, USA

**Abstract**

Terahertz (THz) spin waves or their quanta, magnons, can be efficiently excited by acoustic phonons because these excitations have similar wavevectors in the THz regime. THz acoustic phonons can be produced using photoacoustic phenomena but typically have a low population and thus a relatively low displacement amplitude. The magnetization amplitude and population of the acoustically excited THz magnons are thus usually small. Using analytical calculations and dynamical phase-field simulations, we show that a freestanding metal/magnetic-insulator (MI)/dielectric multilayer can be designed to produce large-amplitude THz spin wave via cavity-enhanced magnon-phonon interaction. The amplitude of the acoustically excited THz spin wave in the freestanding multilayer is predicted to be more than ten times larger than in a substrate-supported multilayer. Acoustically excited nonlinear magnon-magnon interaction is demonstrated in the freestanding multilayer. The simulations also indicate that the magnon modes can be detected by probing the charge current in the metal layer generated via spin-charge conversion across the MI/metal interface and the resulting THz radiation. Applications of the freestanding multilayer in THz optoelectronic transduction are computationally demonstrated.

[*]E-mail: jhu238@wisc.edu

## I. Introduction

One challenge in the field of magnonics is to generate coherent magnons, quanta of spin waves with distinct wavelengths and phases, in the terahertz (THz) frequency range [1]. The generation of coherent THz spin waves has the potential to enable wave-based computing circuits [2] with orders of magnitude higher operation speed than existing gigahertz (GHz) technologies. THz spin waves, due to their nanometer (nm) scale wavelength, are dominated by Heisenberg exchange interactions and therefore called exchange magnons [3]. Strategies for exciting such nm-scale wavelength THz exchange magnons include establishing an external stimulus with a temporal frequency spectrum that overlaps the target THz frequencies.

Experimentally, the excitation of coherent THz exchange magnons by a pulsed THz spin current has been demonstrated in ferromagnetic metal thin films via interfacial spin transfer torque [4,5] or spin-orbit torque [6]. In parallel, a bulk excitation approach is to have THz acoustic phonons propagating inside the magnet and inducing coherent magnetization oscillation via magnon-phonon interaction [7]. Since the wavenumber of acoustic phonons ($k_{ph}$) is similar to the wavenumber of exchange magnons ($k_m$) in the THz regime [8], the magnon-phonon coupling strength, which is proportional to $(k_{ph}k_m)^{1/2}$, can be high [9] and can lead to highly efficient magnon excitation. Experimental efforts have been pursued based on Al(back electrode)/GaAs(substrate)/(Ga,Mn)As(film) heterostructure [10,11], in which a THz acoustic pulse is generated by femtosecond (fs)-duration optical excitation of the Al film. The acoustic pulse propagates across the GaAs substrate and induces magnetic excitation in the (Ga,Mn)As film. The frequency window of the acoustic pulse reaches up to ~0.15 THz in Ref. [10] and 0.3 THz in Ref. [11]. Although magnon modes in the same frequency range should in principle be excited, the frequencies of the experimentally observed magnon modes were below 30 GHz. It appears possible that displacement amplitude of the THz acoustic phonon modes is small in the photoinduced acoustic pulse and that, as a result, the amplitudes of the THz magnon modes could be too small to be detected using the methods reported in Refs. [10] and [11]. Moreover, the optical penetration depth in the time-resolved magneto-optical Kerr effect (MOKE) measurements used in [10,11] exceeds the nm-scale wavelength of THz spin waves. The MOKE signal is thus likely reduced because of spatial averaging over multiple spin wave wavelengths. In this article, we computationally demonstrate that these limitations in magnon excitation and detection can be overcome by employing a hybrid magnon-photon cavity that leverages cavity-enhanced resonant magnon-phonon interaction and spin-charge conversion. Moreover, the simulations indicate that the hybrid magnon-phonon cavity structure can enable the conversion of a fs optical pulse to a nanosecond (ns) THz electric current pulse, which can potentially be exploited to achieve high-quality-factor THz optoelectronic transduction for high-data-rate wireless communication.

The proposed hybrid magnon-phonon cavity consists of a freestanding metal/magnetic-insulator(MI)/dielectric multilayer, as shown in Fig. 1(a). This structure specifically enables the excitation of large-amplitude THz standing exchange spin waves via long-duration, spatially extended, resonant magnon-phonon interaction. The paramagnetic metal layer serves both as a photoacoustic transducer for generating the THz acoustic phonons via fs laser irradiation and a spin-charge current transducer to enable the electrical detection of the acoustically excited THz magnons. Compared to all-metallic THz cavity structures, including freestanding Ni single-layer [12,13], Ni/Au bilayer and Au/Ni/Au multilayer [14], the proposed metal/MI/dielectric multilayer should have smaller eddy current loss, and therefore is more suitable for high-frequency applications. From a fundamental perspective, we computationally reveal the principle of resonant magnon-phonon interaction in the freestanding multilayer using a recently developed dynamical phase-field model [8,15,16] that incorporates nonlinear magnetoacoustic dynamics (i.e., nonlinear relations between the excited spin wave amplitude and the driving acoustic strain), which was omitted in previously existing theoretical studies [12–14,17,18]. Incorporating such nonlinearity not only enables a more accurate calculation of the magnon dynamics and population, but also allow us to model and design novel nonlinear magnonic devices (e.g., magnon-based recurrent neural network [19] and nonlinear switch [20]) with acoustic drive.

The phase-field simulation results show that the proposed freestanding multilayer can enable frequency-selective excitation of coherent exchange magnon modes in the 0.1-1 THz range. Comparative simulations show that the predicted amplitude of THz standing spin wave in the freestanding multilayer is more than ten times larger than in a substrate-supported multilayer. For magnon detection, the simulations show that both the time-dependent electric current in the metal layer and the resultant free-space electromagnetic radiation retain the spectral information of the acoustically excited THz magnon population. The simulation results also indicate that both the amplitude and frequency of the acoustically excited THz spin waves, and the resulting THz electric current, can be tuned by varying a bias magnetic field. The quality factor of a THz optoelectronic transducer based on the present freestanding multilayer is computationally evaluated.

## II. Design Rationale

The freestanding metal/MI/dielectric multilayer, shown in Fig. 1(a), enables the formation of multiple harmonic modes of standing-wave acoustic phonons and magnons. The modes describing the angular wavenumbers of the phonon and magnon are labeled by integer values ($n$, $m$=0, 1, 2, …∞), as shown in Fig. 1(b). The paramagnetic metal layer allows for (i) converting the incident fs optical pulse into a picosecond (ps) acoustic pulse via electron-phonon coupling and thermal expansion [21–24] and (ii) converting the longitudinal spin current to a transverse charge current via the inverse spin Hall effect (iSHE) [25]. The excited magnon modes can be detected by measuring the THz current pulse in the metal (via a coplanar probe tip [26], for frequencies up to 100 GHz) or the free-space electromagnetic (EM) radiation generated by the THz current pulse (via electro-optical sampling) [27].

There are three main principles for selecting the composition of the MI component of the multilayer. First, the MI needs to have a sufficiently large magnetoelastic coupling to enable high magnon-phonon coupling strength [9]. Second, the MI needs to have a high effective spin-mixing conductance to inject large spin current into the adjacent metal layer. Third, the MI layer needs to have a low effective magnetic damping coefficient, which is the sum of the Gilbert damping and other extrinsic contributions such as two-magnon scattering and the spin pumping at the MI/metal interface [28]. A low effective magnetic damping yields a larger magnetization amplitude and a longer lifetime for the magnon. The composition of the metal layer is selected to ensure an efficient photoacoustic transduction and a spin-to-charge conversion for magnon detection. The role of the dielectric layer in the multilayer stack is to provide additional mechanical support for the metal/MI bilayer. Direct integration of the metal/MI bilayer onto a patterned substrate, which enables concentrating a larger portion of the standing acoustic wave in the MI layer, would otherwise be preferred. All layers are best to have low elastic damping to extend the lifetime of the driving acoustic phonons.

It is challenging to find MI and metal materials that simultaneously meet these requirements. Here, we use a freestanding Pt/(001)MgAl$_{0.5}$Fe$_{1.5}$O$_4$/SiN multilayer to illustrate the physical principles. The (001) MgAl$_{0.5}$Fe$_{1.5}$O$_4$ (MAFO) layer is selected because it has a relatively large magnetoelastic coupling coefficient ($B_1$=1.2 MJ m$^{-3}$) and a low Gilbert damping ($\alpha^0$=0.0015 for uniform spin precession, or the $m$=0 mode magnon) at room temperature [29]. Additionally, a large spin Hall angle ($\theta_{Pt}$~0.83) has previously been measured at room temperature for the Pt in a MAFO/Pt bilayer [30], which will result in a high-efficiency spin-to-charge conversion. Other promising MI materials include: (i) YIG, which has an ultra-low Gilbert damping ($\alpha^0$~8×10$^{-5}$ [31]) but weak magnetoelastic coupling ($B_1$=0.3 MJ m$^{-3}$, $B_2$=0.55 MJ m$^{-3}$ [16]) at room temperature; (ii) the rare-earth iron garnet Tb$_3$Fe$_5$O$_{12}$ (TbIG) which has high magnetoelastic coupling ($B_1$=-283.32 MJ m$^{-3}$, $B_2$=-499.815 MJ m$^{-3}$) at cryogenic temperature (4.2 K) [32] and a moderate room-temperature Gilbert damping ($\alpha^0$=0.01-0.02, extrapolated from measurements of similar Tm$_3$Fe$_5$O$_{12}$ films and Tm$_3$Fe$_5$O$_{12}$/Pt bilayers [33]).

### III. Analytical calculations

By solving a set of linearized elastodynamic equations under appropriate boundary conditions (see Appendix A), we derive an analytical formula for the frequencies of the standing acoustic phonon modes in a freestanding Pt/MAFO/SiN multilayer as a function of the thickness of the individual layers. The formula is,

$$\begin{aligned}
&c_{Pt} c_{SiN} \left(-1+e^{\frac{2id_{Pt}\omega_n}{v_{Pt}}}\right)\left(-1+e^{\frac{2id_{MAFO}\omega_n}{v_{MAFO}}}\right)\left(-1+e^{\frac{2id_{SiN}\omega_n}{v_{SiN}}}\right) v_{MAFO}^2 \\
&+c_{MAFO}^2 \left(1+e^{\frac{2id_{Pt}\omega_n}{v_{Pt}}}\right)\left(-1+e^{\frac{2id_{MAFO}\omega_n}{v_{MAFO}}}\right)\left(1+e^{\frac{2id_{SiN}\omega_n}{v_{SiN}}}\right) v_{Pt} v_{SiN} \\
&+c_{MAFO} c_{SiN} \left(1+e^{\frac{2id_{Pt}\omega_n}{v_{Pt}}}\right)\left(1+e^{\frac{2id_{MAFO}\omega_n}{v_{MAFO}}}\right)\left(-1+e^{\frac{2id_{SiN}\omega_n}{v_{SiN}}}\right) v_{Pt} v_{MAFO} \\
&+c_{Pt} c_{MAFO} \left(-1+e^{\frac{2id_{Pt}\omega_n}{v_{Pt}}}\right)\left(1+e^{\frac{2id_{MAFO}\omega_n}{v_{MAFO}}}\right)\left(1+e^{\frac{2id_{SiN}\omega_n}{v_{SiN}}}\right) v_{MAFO} v_{SiN} = 0,
\end{aligned} \qquad (1)$$

where $d_{\text{mater}}$, $c_{\text{mater}}$, and $v_{\text{mater}}$ (mater = Pt, MAFO, SiN) refer to the thickness, elastic stiffness component $c_{11}$, and longitudinal sound speed of the individual layer, respectively and $\omega_n$ is the angular frequency of the standing-wave acoustic phonon modes, with $n = 1, 2, 3, \ldots \infty$. A list of symbols for the main physical quantities used in this paper is provided in Supplemental Material [34]. The first nonzero nontrivial solution of Eq. (1) yields the angular frequency value of the $n$=1 acoustic phonon mode ($\omega_{n=1}$), and so forth for the higher-order modes. Equation (1) can be applied to a freestanding bilayer (single layer) by setting a zero thickness for one (two) layers. When the thickness of one layer is set to zero, Eq. (1) is reduced to an expression for bilayer that is equivalent to that provided in [14].

Previously, Zhuang and Hu derived the dispersion relation of the exchange magnons where the initial equilibrium magnetization vector $\mathbf{m}^0$ can align along any directions, i.e., $f = \frac{\gamma}{2\pi}\sqrt{D^2 k_m^4 + \Omega D k_m^2 - \Lambda}$ [15]. For standing wave magnon modes, the angular wavenumber $k_m = m\pi/d_{\text{MAFO}}$ ($m$=0, 1, 2, ... ∞) depends on the MAFO layer thickness $d_{\text{MAFO}}$. Here the exchange stiffness $D = \frac{2A_{\text{ex}}}{\mu_0 M_s}$ where $A_{\text{ex}}$ is the exchange coupling coefficient and $M_s$ is the saturation magnetization; $\gamma$ is gyromagnetic ratio and $\mu_0$ is vacuum permeability; and $\Omega$ and $\Lambda$ are functions of the initial equilibrium magnetization vector $\mathbf{m}^0$ and bias magnetic field $\mathbf{H}^{\text{bias}}$. Detailed formulae for $\Omega$ and $\Lambda$ are in Appendix B. In this study, we consider the case in which $\mathbf{m}^0$ is 45° off the [110] direction ($\theta=\varphi=45°$, see inset of Fig. 2), which maximizes the torque from the effective magnetoelastic field [8]. Figure 2 shows the frequencies of the standing acoustic phonons, $m$=1, and $m$=2 mode standing magnon as a function of the $d_{\text{MAFO}}$, where the $d_{\text{Pt}}$ and $d_{\text{SiN}}$ are fixed at 6.6 nm and 31.5 nm, respectively. When the frequency of the $m$=1 (or $m$=2) magnon mode is equal to the frequency of the driving acoustic phonon mode, as indicated by the circles, the magnon mode can be resonantly excited. Since the fs-laser-induced acoustic pulse has a broad temporal frequency spectrum, the excitation of $m$=2 magnon mode would occur simultaneously with the excitation of both $m$=1 and $m$=0 magnon modes. For simplicity, we will focus on resonant, selective excitation of $m$=1 magnon mode by keeping the frequency window of the injected acoustic pulse below the frequency of the $m$=2 magnon mode [15].

## IV. Dynamical Phase-Field simulations

To simulate the acoustic excitation of magnons and the resulting spin-charge conversion in the freestanding metal/MI/dielectric multilayer in a coupled fashion, we employ the recently developed dynamical phase-field model that incorporates the coupled dynamics of acoustic phonons, magnons, photons, and plasmons [8,15,16]. The photons (EM waves) result from both the precessing magnetization in the MI (via the magnetic dipole radiation) and the oscillating charge current density in the metal (via the electric dipole radiation). The magnetic-field component of the EM waves will affect — albeit not significantly in this case — the magnetization dynamics in the MI layer. Furthermore, since the emitted EM wave can induce large eddy current density ($J^p$) in the metal layer, the $\mathbf{J}^c$ in the metal layer is a sum of both the $\mathbf{J}^{iSHE}$ (the contribution from the spin-charge conversion via the iSHE) and the $\mathbf{J}^p$. By coupling the dynamics of plasmons in the metal with the photon dynamics, both the $\mathbf{J}^c$ and the EM wave can be accurately simulated [15].

In our dynamical phase-field model, the Landau-Lifshitz-Gilbert (LLG) equation is used to describe the temporal evolution of the normalized magnetization $\mathbf{m}=\mathbf{M}/M_s$ in the MI layer,

$$\frac{\partial \mathbf{m}}{\partial t} = -\frac{\gamma}{1+\alpha^2} \mathbf{m} \times \mathbf{H}^{\text{eff}} - \frac{\alpha\gamma}{1+\alpha^2} \mathbf{m} \times (\mathbf{m} \times \mathbf{H}^{\text{eff}}), \tag{2}$$

where $\alpha$ is the effective magnetic damping coefficient of the MI layer. In the proposed structure, $\alpha=\alpha^0+\alpha^s$ where $\alpha^0$ is the Gilbert damping coefficient for the $m=0$ mode magnon, and $\alpha^s = \frac{g\mu_B}{4\pi M_s} G_{\text{eff}}^{\uparrow\downarrow} \frac{1}{d}$ [35] describes the magnetic damping induced by spin pumping from the metal/MI interface into the paramagnetic metal. Here $g = 2.05$ is the $g$-factor [29], $\mu_B$ is the Bohr magneton; $G_{\text{eff}}^{\uparrow\downarrow}$ is the real part of the effective spin-mixing conductance. The total effective magnetic field $\mathbf{H}^{\text{eff}}=\mathbf{H}^{\text{anis}}+\mathbf{H}^{\text{exch}}+\mathbf{H}^{\text{dip}}+\mathbf{H}^{\text{bias}}+\mathbf{H}^{\text{mel}}+\mathbf{H}^{\text{EM}}$ is the sum of the magnetocrystalline anisotropy field $\mathbf{H}^{\text{anis}}$, the magnetic exchange coupling field $\mathbf{H}^{\text{exch}}$, the magnetic dipolar coupling field $\mathbf{H}^{\text{dip}}$, the bias magnetic field $\mathbf{H}^{\text{bias}}$, the magnetoelastic field $\mathbf{H}^{\text{mel}}$ and the magnetic field component $\mathbf{H}^{\text{EM}}$ of the EM wave. Expressions for $\mathbf{H}^{\text{anis}}$ and the $\mathbf{H}^{\text{dip}}$, which are both are a function of $\mathbf{m}$, and the $\mathbf{H}^{\text{exch}}$, a function of $\nabla^2 \mathbf{m}$, can be found in [15]. The $\mathbf{H}^{\text{bias}}$ is applied to stabilize the initial equilibrium magnetization $\mathbf{m}^0$ to be 45° off the [110] direction, as mentioned in Section III. $\mathbf{H}^{\text{mel}}$ is a function of $\mathbf{m}$ and local strain $\boldsymbol{\varepsilon}$, given by [15],

$$H_i^{\text{mel}} = -\frac{2}{\mu_0 M_s}[B_1 m_i \varepsilon_{ii} + B_2(m_j \varepsilon_{ij} + m_k \varepsilon_{ik})], i,j=x,y,z; j \neq i, k \neq i,j, \tag{3}$$

It is noteworthy that the LLG equation (Eq. (2)) is intrinsically nonlinear, e.g., the frequency of magnetic excitation can double the frequency of the driving effective field when the amplitude of the magnetic excitation is sufficiently large. Such nonlinearity can be analytically understood based on the conservation of magnetization vector length (i.e., $m_x^2+m_y^2+m_z^2=1$) and the method of successive approximation (see details in [36]).

The local strain $\boldsymbol{\varepsilon}$ is $\varepsilon_{ij}=\frac{1}{2}\left(\frac{\partial u_i}{\partial j}+\frac{\partial u_j}{\partial i}\right)$ ($i,j = x, y, z$). The evolution of the mechanical displacement $\mathbf{u}$ is governed by the elastodynamics equation,

$$\rho\frac{\partial^2 \mathbf{u}}{\partial t^2} = \nabla \cdot (\boldsymbol{\sigma}+\beta\frac{\partial \boldsymbol{\sigma}}{\partial t}), \tag{4}$$

where stress $\boldsymbol{\sigma} = \mathbf{c}(\boldsymbol{\varepsilon}-\boldsymbol{\varepsilon}^0)$; $\rho$, $\beta$ and $\mathbf{c}$ are the mass density, stiffness damping coefficient and elastic stiffness coefficient, respectively. The stress-free magnetostrictive strain $\boldsymbol{\varepsilon}^0$ is $\varepsilon_{ii}^0=\frac{3}{2}\lambda_{100}^M\left(m_i^2-\frac{1}{3}\right)$, $\varepsilon_{ij}^0=\frac{3}{2}\lambda_{111}^M m_i m_j$, with $i,j=x,y,z$, where $\lambda_{100}^M$ and $\lambda_{111}^M$ are the magnetostrictive coefficients of the MI layer (assuming a cubic symmetry). To model the injection of fs-laser-induced ps-duration acoustic pulse, a Gaussian-shaped stress pulse $\sigma_{zz}(z = d_{\text{MAFO}}+d_{\text{Pt}}, t) = \sigma_{\max} \exp[-\frac{t^2}{2\tau^2}]$ is applied at the top surface of the Pt layer, where $\sigma_{\max}$ is the peak magnitude of the applied stress and $\tau$ is a free parameter that controls the pulse duration. Varying these two parameters allows us to tune the peak amplitude and frequency window of the acoustic pulse injected into the Pt. Based on existing experiments, the frequency window of such fs-laser-induced acoustic pulse is typically in the sub-THz range [10,11] but covers up to nearly 3 THz [37]; the peak strain amplitude is typically in the order of $10^{-3}$ [22,23,37] but can reach 1% [38].

The EM wave is described by Maxwell's equations. The two governing equations for the magnetic and electric field components are,

$$\nabla \times \mathbf{E}^{\text{EM}} = -\mu_0 \left( \frac{\partial \mathbf{H}^{\text{EM}}}{\partial t} + \frac{\partial \mathbf{M}}{\partial t} \right), \tag{5}$$

$$\nabla \times \mathbf{H}^{\text{EM}} = \varepsilon_0 \varepsilon_r \frac{\partial \mathbf{E}^{\text{EM}}}{\partial t} + \mathbf{J}^{\text{f}} + \mathbf{J}^{\text{p}}, \tag{6}$$

where $\mathbf{E}^{\text{EM}}$ is the electric field component of the EM wave; $\mathbf{M} = M_s \mathbf{m}$ is the local magnetization in the MI layer, and $\mathbf{m}$ is obtained by solving the LLG equation (Eq. (2)); $\varepsilon_0$ and $\varepsilon_r$ are vacuum and relative permittivity, respectively; $\mathbf{J}^{\text{f}}$ and $\mathbf{J}^{\text{p}}$ are the free current density and polarization current density, respectively, in the metal. $\mathbf{J}^{\text{p}}$ is induced by the electric field $\mathbf{E}^{\text{EM}}$ in dispersive medium such as the metallic Pt, which leads to the absorption and reflection of the EM wave. $\mathbf{J}^{\text{p}}$ is obtained by solving a time-dependent auxiliary differential equation based on the Drude model,

$$\frac{\partial \mathbf{J}^{\text{p}}}{\partial t} + \frac{\mathbf{J}^{\text{p}}}{\tau_e} = \varepsilon_0 \omega_p^2 \mathbf{E}^{\text{EM}}, \tag{7}$$

where $\omega_p$ and $\tau_e$ are the plasma frequency and electron relaxation time in the metal, respectively.

These coupled equations of motions for $\mathbf{M}$, $\mathbf{u}$, EM fields, and the $\mathbf{J}^{\text{p}}$ are numerically solved in a one-dimensional (1D) simulation system consisting of Pt/MAFO/SiN with free space above and below the multilayer. The physical quantities describing the system vary only along the $z$ axis. To demonstrate that the use of a 1D system is a reasonable approximation, we built a 2D multiphase system as shown in Fig. 1(a) and simulated the evolution of the distributions of the acoustic phonons and magnons after the acoustic pulse injection. It is found that both the acoustic phonons and magnons have the same amplitude and phase within the $xy$ plane, with nonuniformity arising only near the lateral surfaces of the Pt and MAFO, as shown in Appendix C.

Since the local magnetization $\mathbf{m}$ is spatially uniform in the $xy$ plane, the free current density $\mathbf{J}^{\text{f}}$ in the Pt layer, which is converted from the spin current density $\mathbf{J}^{\text{s}}$ via the iSHE ($\mathbf{J}^{\text{f}} = \mathbf{J}^{\text{iSHE}}$), is also spatially uniform in the $xy$ plane. The spin current density $\mathbf{J}^{s,0}(t) = \mathbf{J}^s(z=d_{\text{MAFO}}, t)$ at the MAFO/Pt interface is calculated via the relation [39],

$$\mathbf{e}_n \cdot \mathbf{J}^{s,0} = \frac{\hbar}{4\pi} G_{\text{eff}}^{\uparrow\downarrow} \left( \mathbf{m} \times \frac{\partial \mathbf{m}}{\partial t} \right), \tag{8}$$

where $\mathbf{e}_n$ is the unit vector normal to the MAFO/Pt interface and points to Pt, i.e., $\mathbf{e}_n$=[0,0,1]; $\hbar$ is the reduced Planck constant. Note that $\mathbf{J}^s = J^s_{ij}$ is a second-rank tensor, where the subscript '$i$' refers to the direction of the spin current flow and '$j$' denotes the direction of the spin polarization [40]. Thus, $\mathbf{e}_n \cdot \mathbf{J}^{s,0} = J^{s,0}_{zj}$. The spin current density in the Pt layer $J^s_{zj}(z,t)$ decays as a function of the distance from the MAFO/Pt interface ($z=d_{\text{MAFO}}$), $J^s_{zj}(z,t) = J^{s,0}_{zj}(t) \frac{\sinh[(d_{\text{MAFO}}+d_{\text{Pt}}-z)/\lambda_{\text{sd}}]}{\sinh(d_{\text{Pt}}/\lambda_{\text{sd}})}$, which is obtained by solving the 1D spin diffusion equation under the boundary conditions of $J^s_{zj}(z=d_{\text{MAFO}},t) = J^{s,0}_{zj}(t)$ and $J^s_{zj}(z=d_{\text{MAFO}}+d_{\text{Pt}},t) = 0$ [35,41], with $d_{\text{Pt}}$ indicating the thickness of the Pt layer and $\lambda_{\text{sd}}$ is the spin diffusion length in Pt. The $\mathbf{J}^{\text{iSHE}}$ in the Pt layer is calculated based on $\mathbf{J}^{\text{iSHE}}(z,t) = \theta_{\text{Pt}} \frac{2e}{\hbar} \mathbf{e}_n \times \mathbf{J}^s_{zj}(z,t)$, where $\theta_{\text{Pt}}$ is the spin Hall angle of Pt and $e$ is the elementary charge. Specifically, $J^{\text{iSHE}}_x(z,t) = -\theta_{\text{Pt}} \frac{2e}{\hbar} J^s_{zy}(z,t)$, $J^{\text{iSHE}}_y(z,t) = \theta_{\text{Pt}} \frac{2e}{\hbar} J^s_{zx}(z,t)$, and $J^{\text{iSHE}}_z(z,t) = 0$. Equation (8) and the formulae of $\mathbf{J}^{\text{iSHE}}(z,t)$ together indicate that magnons of higher frequency but smaller amplitude of magnetization variation can still induce a larger peak spin current density and the resulting iSHE charge current density (see detailed analysis in Appendix D). For the MAFO/Pt interface [30], the effective spin-mixing conductance $G^{\uparrow\downarrow}_{\text{eff}}$= 3.36×10$^{18}$ m$^{-2}$, the spin diffusion length in Pt $\lambda_{\text{sd}}$ = 3.3 nm, and the spin Hall angle of Pt $\theta_{\text{Pt}}$ = 0.83.

When irradiating the Pt film with a fs laser pulse, the temperature gradient across the MAFO/Pt interface can also lead to the injection of spin current into the Pt via the interfacial spin Seebeck effect [42–45]. However, such thermally pumped spin current typically persists for at most tens of ps after the fs laser excitation [46], which is one-to-two order of magnitude shorter than the lifetime of spin current from acoustic spin pumping (10$^{-10}$-10$^{-9}$ s, as will be shown later). Therefore, thermal spin pumping is not modeled herein for simplicity. The other material parameters used for simulations are listed in Appendix E. The numerical methods for solving the LLG, elastodynamic, and Maxwell's equations are described in Appendix F.

**V. Resonant acoustic excitation of THz magnons by dynamical phase-field simulations**

Figure 3(a) shows the evolution of the strain $\varepsilon_{zz}(t)$ at the MAFO/Pt interface of freestanding Pt(6.64 nm)/MAFO(12.45 nm)/SiN(31.54 nm) multilayer and a substrate-supported multilayer of Pt(6.64 nm)/MAFO(12.45 nm)/SiN. In the latter structure, the absorbing boundary condition is applied on the bottom surface of SiN to make it a perfect acoustic sink (see details in Appendix F). The two parameters determining $\sigma_{zz}(t)$ (see Section IV) were selected to be $\sigma_{\max}$=3 GPa and $\tau$ =1.5 ps, which lead to an acoustic spectrum covering up to 300 GHz and a peak strain pulse $\varepsilon_{zz}(t)$ amplitude of 0.85% in the Pt. As shown in Fig. 3(a), $\varepsilon_{zz}(t)$ consists of multiple cycles of acoustic oscillation that persist for more than 1 ns in the freestanding multilayer. The acoustic frequency spectrum (Fig. 3(b)) displays three peaks at 67 GHz, 154 GHz, and 240 GHz, which agree well with the analytically calculated frequencies of the $n$=1, 2, 3 acoustic phonon modes, respectively. In contrast, $\varepsilon_{zz}(t)$ exhibits a single-cycle of oscillation that persists for only ~9 ps in the substrate-supported Pt/MAFO/SiN multilayer, and there is no peak in its frequency spectrum.

Figure 3(c) shows the numerically simulated change in magnetization $\Delta m_x(t)$ at the MAFO/Pt interface of the freestanding multilayer. Its frequency spectrum, shown in Fig. 3(d), indicates the presence of three magnon modes at 67 GHz, 154 GHz, and 240 GHz that are induced by the $n=1$, $n=2$, and $n=3$ acoustic phonon modes, respectively. Remarkably, the peak at 154 GHz has a larger spectral amplitude (i.e., larger magnon population) than the peak at 67 GHz, even though the driving 154 GHz ($n=2$) phonon mode has a smaller spectral amplitude than the 67 GHz ($n=1$) phonon (c.f., Fig. 3(b)). The large response at 154 GHz clearly indicates the resonant interaction between the driving $n=2$ phonon mode and the $m=1$ magnon mode. In contrast, for the substrate-supported Pt/MAFO/SiN multilayer, the temporal profile of $\Delta m_x(t)$ is dominated by a low-frequency variation corresponding to the $m=0$ magnon mode (the ferromagnetic resonance, or FMR). The simulated FMR frequency is ~0.5 GHz, see Fig. 3(d) inset, which agrees well with the analytically calculated value 0.53 GHz. The 154 GHz $m=1$ mode magnon, in this case, adds to the profile of $\Delta m_x(t)$ in the form of a small-amplitude, high-frequency oscillation (see Fig. 3(c)). To extract the real-space amplitude of magnetization variation for the $m=1$ mode magnon in both the freestanding and substrate-supported multilayer, we performed inverse Fourier transform of the 154 GHz peaks. As shown in Appendix G, the amplitude of magnetization variation for the $m=1$ magnon mode in the MAFO has a maximum peak amplitude of $\Delta m_x \sim 0.082$ in the freestanding multilayer, 12 times larger than that in the substrate-supported multilayer.

For the freestanding multilayer, Figure 3(d) also indicates the presence of multiple magnon modes that do not have counterparts in the frequency spectrum of acoustic phonons. These additional magnon modes are attributed to the acoustically excited nonlinear magnon-magnon interaction. Specifically, the effect of each standing-wave acoustic phonon mode on magnetic excitation is equivalent to the effect of applying a sinusoidal magnetic field of a distinct frequency $f_n$ ($n=1,2,3...$). When the magnetization amplitudes of the acoustically excited magnon modes are sufficiently large, the nonlinearity of the LLG equation [36] would lead to nonlinear effects such as frequency mixing ($f=n_1 f_1 \pm n_2 f_2 \pm ...$, $n_{1,2}=0,1,2...$) and frequency doubling ($f=cf_n$, $c=1,2,3...$). As shown in Fig. 3(d), there are three magnon modes arising from frequency mixing, all involving the resonantly excited $m=1$ magnon mode ($f_2=f_{m=1}$) that has a large magnetization amplitude. There is one magnon mode resulting from the second harmonic generation of the $f_1$ magnon mode. Notably, the $f_1+f_2$ magnon mode, induced through the mixing of two dominant magnon modes, has a larger spectral amplitude than the linearly excited $f_3$ ($m=3$) magnon mode.

Figure 3(e) shows the evolution of the total charge current density $\mathbf{J}^c(z=d_{MAFO}, t)$ at the Pt/MAFO interface in the freestanding multilayer, which is a sum of the iSHE charge current $\mathbf{J}^{iSHE}(z=d_{MAFO}, t)$ and the polarization current $\mathbf{J}^p(z=d_{MAFO}, t)$. The $\mathbf{J}^p$ has a similar temporal profile to that of the $\mathbf{J}^{iSHE}$, but it has a smaller amplitude and is 180° out of phase (see Appendix H). Similarly to the profile of the $\Delta m_x(t)$ in Fig. 3(c), the total current density $J_x^{tot}(t)$ contains a mixture of low-frequency and high-frequency components and persists for several ns. The two dominant frequencies (67 GHz and 154 GHz) in the frequency spectrum of the current (Fig. 3(f)) are the same as those in the spectrum of the $\Delta m_x(t)$. As shown in Fig. 3(f), the spectral amplitude of the 154 GHz peak is about 10 times larger than the 67 GHz peak. This enhancement is even more significant than that shown in Fig. 3(d), because the 154 GHz magnon leads to a larger $\mathbf{J}^{\text{iSHE}}$ via spin pumping (see Eq. (8)) as compared to the 67 GHz magnon. The spectral information of all nonlinearly excited magnon modes is also retained in the frequency spectrum of $J_x^c(t)$. By comparison, $J_x^c(t)$ in the substrate-supported Pt/MAFO/SiN multilayer has one order of magnitude smaller amplitude and a much shorter lifetime. The frequency spectrum of the $J_x^c(t)$ shows a single peak at 154 GHz which corresponds to the $m=1$ magnon mode. There are no peaks corresponding to nonlinear THz magnon modes due to the relatively small magnetization amplitudes of the acoustically excited THz magnons. The temporal profile of the electric field component $\mathbf{E}^{\text{EM}}(t)$ of the EM wave emitted into free space, as shown in Appendix I, is similar to that of the $J_x^c(t)$. The peak amplitude of $\mathbf{E}^{\text{EM}}$ is ~400 V/m in the freestanding multilayer and ~36 V/m in the substrate-supported multilayer. These peak electric fields are large enough for measurement by free-space electro-optical sampling [47] and their frequencies (154 GHz and 67 GHz) are also within the detectable range [27]. Alternatively, it may be possible to directly measure the charge current in this frequency range in the Pt layer using a coplanar probe tip followed by *fs* laser irradiation (known as ultrafast iSHE measurement [26]).

It is possible to dynamically tune the frequencies and temporal profile of the $\mathbf{J}^c$ in the Pt layer and hence the free-space EM radiation by varying the bias magnetic field. Figure 4(a) shows the analytically calculated frequencies (*f*) of the $m=0$ and $m=1$ magnon modes as a function of the magnitude of the bias magnetic field $|\mathbf{H}^{\text{bias}}|$ in the freestanding Pt(6.64)/MAFO(12.45)/SiN(31.54) multilayer. The direction of the $\mathbf{H}^{\text{bias}}$ is varied to keep the equilibrium magnetization 45° off the [110] direction ($\theta=\varphi=45°$). The frequencies of the standing acoustic phonon modes, which do not vary with the magnetic field, are plotted as horizontal lines. At $|\mathbf{H}^{\text{bias}}|=0.087$ T (as in Fig. 3), the $m=1$ magnon mode has a frequency of 154 GHz, which leads to resonant interaction with the $n=2$ phonon mode. At $|\mathbf{H}^{\text{bias}}|=3.02$ T, the $m=1$ magnon mode shows an increased frequency to 240 GHz, and hence can resonantly interact with the $n=3$ phonon mode, and so forth.

To demonstrate resonant excitation of the $m=1$ magnon mode by the higher-order $n=3$ phonon mode, we inject a strain pulse with a frequency window reaching 500 GHz by setting the parameter $\tau$ in the applied stress $\sigma_{zz}(t)$ to be 0.8 ps and keeping the $\sigma_{\text{max}}=3$ GPa the same as that in Fig. 3. The acoustically excited magnon dynamics and charge current dynamics are then simulated under $|\mathbf{H}^{\text{bias}}|=3.02$ T. Figure 4(b) shows the evolution of the injected $\varepsilon_{zz}(t)$ at the MAFO/Pt interface. As shown by its frequency spectrum in Fig. 4(c), phonon modes $n=1$ to 5 can all be excited, and their frequencies agree well with the analytical calculations via Eq. (1). The evolution of the total current density $J_x^c(t)$ at the MAFO/Pt interface is shown in Fig. 4(d). The duration of $J_x^c(t)$ is shorter than that the case of 154 GHz (c.f., Fig. 3(e)) due to the shorter lifetime of the driving 240 GHz phonon. Remarkably, as shown in Fig. 4e, the spectral amplitude of the 240 GHz magnon peak is the largest even though the 240 GHz phonon peak has a negligibly small spectral amplitude (*c.f.*, Fig. 4(c)). The other two frequency peaks in Fig. 4(e) are contributed by the non-resonant $m$=0 magnon (FMR) mode (86 GHz) and the non-exchange magnon mode (67 GHz) induced by the $n$=1 phonon mode. Likewise, nonlinear magnon modes resulting from frequency mixing or doubling are also induced (see Appendix J).

Due to this capability of converting a fs laser pulse to a ns-lifetime charge current with oscillation frequencies over 100 GHz, the proposed freestanding multilayer can potentially be utilized to develop a dynamically tunable, on-chip THz optoelectronic transducer, which can further be used to design an on-chip THz optoelectronic oscillator (OEO) for next-generation wireless communication application. The maximum achievable operation frequency of existing OEO systems is usually below 100 GHz [48–50], for which one key limitation is the relatively low operation frequency of the optoelectronic transducer (typically a semiconductor-based photodetector) [48,49,51]. There are other approaches that permit converting a fs laser pulse to a THz charge current pulse, including photoconductive (Auston) switch [52,53] and spintronic THz emitter (STE) [54–57], but the THz current pulses are broadband and typically persist for at most a few ps. For comparison, we evaluated the frequency ($f$) dependence of the quality factor ($Q=f/\Delta f$) of the freestanding multilayer. First, we perform analytical calculations (similarly to Fig. 2) to identify the $d_{MAFO}$ and $d_{SiN}$ (see Table 1 in Appendix K for values) that enable resonant excitation of $m$=1 magnon mode by acoustic phonons of different modes within the 0.03-1 THz range. $d_{Pt}$ is fixed at 6.6 nm. Second, dynamical phase-field simulations are performed to obtain the charge current density at the MAFO/Pt interface for each combination of $d_{MAFO}$ and $d_{SiN}$, which allows for extracting the linewidth $\Delta f$ (full width half maximum) of the target peak frequency. For an oscillatory time-domain signal, a longer lifetime leads to a narrower linewidth $\Delta f$ and hence a larger $Q$. For further comparison, we also evaluated the $Q$-$f$ relation in the substrate-supported Pt/MAFO/SiN(acoustic sink) multilayer, which also displays multi-cycle charge current pulse but with fewer cycles (hence should have a smaller $Q$) and smaller amplitude.

In both the freestanding and substrate-supported multilayer, the lifetime of the charge current pulse is determined largely by the lifetime of the acoustically excited magnons at the MAFO/Pt interface. In the freestanding multilayer that involves extended resonant magnon-phonon interaction, the magnon lifetime is determined by both the elastic damping and the magnetic damping, and both damping terms become more significant at higher frequencies. Specifically, the lifetime of driving acoustic phonon ($\tau_{ph}$), which can be analytically calculated as $\tau_{ph}=\frac{\beta}{\sqrt{1+\omega^2\beta^2}-1}$, is shorter at larger $\omega$, indicating an enhanced elastic damping. Moreover, the spin-pumping-induced damping ($\alpha^s$), which is part of the effective magnetic damping coefficient ($\alpha$), is proportional to $1/d_{MAFO}$. To obtain higher-frequency $m$=1 mode magnon, smaller $d_{MAFO}$ needs to be used (see Fig. 2), resulting in a more significant magnetic damping at higher operation frequencies.

As shown in Fig. 5, the $Q$ of the freestanding multilayer decreases largely linearly with the $f$, with a high $Qf$ product of above 10 THz which is one order-of-magnitude higher than both the Auston switch and STE. To evaluate the role of elastic and magnetic damping, two control simulations were performed for the Pt(6.64 nm)/MAFO(12.45 nm)/SiN(31.54 nm) freestanding multilayer by individually setting $\beta=0$ or $\alpha=0$, with a resonant frequency of about 154 GHz (the same as in Fig. 3). As shown in Appendix L, the $Q$ increases from about 155 to 428 under $\alpha=0$ while increases to 588 under $\beta=0$, suggesting a more substantial role of the elastic damping in the freestanding multilayer. By contrast, in the substrate-supported multilayer, the lifetime of the acoustically excited magnons is largely determined by the magnetic damping alone because the injected acoustic pulse only stays in the MAFO layer briefly (~1.55 ps for a 12.5-nm-thick film). For this reason, $Q$ of the substrate-supported multilayer decreases less drastically with the increasing $f$ compared to the $Q - f$ relation of the freestanding multilayer. Nevertheless, both the $Q$ and the charge current amplitude of the freestanding multilayer, due to the resonant magnon-phonon interaction, are substantially larger than the substrate-supported multilayer even at relatively high frequencies (similarly to those in Fig. 3(e)).

It is also worth comparing the acoustic-to-magnetic energy conversion efficiency of the freestanding and substrate-supported multilayer, which critically determines the overall efficiency of the THz optoelectronic transducer. The areal energy of the injected acoustic wave is evaluated using $F_a = \int_0^{t_0} \left[ -\frac{\partial u_z(z=z_0,t)}{\partial t} K \frac{\partial u_z(t)}{\partial z} \Big|_{z=z_0} \right] dt$ in a single-layer Pt where the Gaussian-shaped stress pulse $\sigma_{zz}(t)$ is applied to its top surface ($z=z_0$) yet the absorbing boundary condition is applied to the bottom surface to avoid acoustic wave reflection. Here $u_z$ is the mechanical displacement obtained from the simulations, $K$ is the bulk modulus of the Pt, and $t_0$ is the duration of the applied stress pulse. For a stress pulse with $\sigma_{max}$=3 GPa and $\tau$=1.5 ps (the same as those used in Fig. 3), $F_a \approx 0.2256$ J/m². The magnetic energy converted from the acoustic wave is evaluated by calculating the magnetic energy dissipation over the entire process of excitation and relaxation, i.e., $F^m = \int_0^{t_1} \int_0^d \frac{\partial f(z,t)}{\partial t} dz\, dt$, where $d$ is the thickness of magnetic layer; $t_1$ is the overall duration of the magnetic excitation; the dissipation rate of the magnetic energy density $\frac{\partial f}{\partial t} = \frac{\partial f}{\partial \mathbf{m}} \frac{\partial \mathbf{m}}{\partial t} = -\mu_0 M_s (\mathbf{H}^{\mathrm{anis}} + \mathbf{H}^{\mathrm{dip}} + \mathbf{H}^{\mathrm{bias}} + \mathbf{H}^{\mathrm{mel}}) \frac{\partial \mathbf{m}}{\partial t}$ [39]. After some algebra and omitting the contribution from $\mathbf{H}^{\mathrm{anis}}$ due to the relatively weak magnetocrystalline anisotropy of the MAFO, one has $F^m \approx \int_0^{t_1} \int_0^{d_{\mathrm{MAFO}}} (\mu_0 M_s^2 \Delta m_z + 2B_1 m_z \varepsilon_{zz}) \frac{\partial m_z}{\partial t} dz\, dt$. Using the simulated $\mathbf{m}(z,t)$ and $\varepsilon_{zz}(z,t)$ in the case of Fig. 3 as the input, $F^m$ is calculated to be 9.26×10⁻⁵ J/m² for the freestanding multilayer and 1.26×10⁻⁷ J/m² for substrate-supported multilayer. Thus, the acoustic-to-magnetic energy conversion efficiency in the freestanding multilayer (~4×10⁻⁴) is approximately three orders of magnitude higher than that in the substrate-supported multilayer (~5.6×10⁻⁷). From the analytical formula of $F^m$, the conversion efficiency in the freestanding multilayer can be further improved by (i) using a MI layer with larger magnetoelastic coupling coefficient $B_1$ and saturation magnetization $M_s$ (such as TbIG [32]) and (ii) optimizing the spatial profile overlap between the magnon and acoustic phonon modes such that a larger portion of $\varepsilon_{zz}(z,t)$ can locate within the MI layer (e.g., using a freestanding metal/MI bilayer) along with the wavenumber matching.

## VI. Conclusions

In summary, we have computationally designed a hybrid magnon-phonon cavity, which consists of a freestanding metal/MI/dielectric multilayer and can enable the frequency-selective excitation of large-amplitude THz spin waves via cavity-enhanced magnon-phonon interaction. We developed an analytical formula to identify the individual layer thickness (Fig. 2) and the bias magnetic field (Fig. 4(a)) that leads to the resonant magnon-phonon interaction. We performed dynamical phase-field simulations to model the spatiotemporal profiles of the acoustically excited magnon modes, the resulting charge current pulse and free-space EM radiation both of which are found to retain the spectral information of both the linearly and nonlinearly excited THz magnon modes. The simulation results suggest that it is possible to excite and detect large-amplitude THz spin wave in the proposed freestanding multilayer by ultrafast iSHE measurement or THz emission spectroscopy with fs optical pump.

From a fundamental perspective, the proposed freestanding multilayer would be a well-suited platform for studying the damping mechanisms of magnons in the THz regime (e.g., the possible inertial effects [58–61]) due to the capability of enabling ns-lifetime THz magnons. Furthermore, the extended magnon-phonon interaction time in the freestanding multilayer makes it possible to realize the hybridization of magnons and phonons in the THz regime by matching both their frequency and wavenumber via geometrical design (see Appendix M).

For applications, the proposed freestanding multilayer can potentially be used to design a THz optoelectronic transducer, due to its unique capability to convert a fs laser pulse into a ns charge current pulse with dominant frequency peak in the 0.1-1 THz range. It has a one-order-of-magnitude higher $Q$ factor than previously existing technologies (Auston switch [52,53], spintronic THz emitter [54–57]) that also involve the use of fs optical pump to generate THz current pulses. Combined with the dynamical tunability, such high-$Q$ magnonic THz optoelectronic transducer provides attractive potential for the THz wireless communication [62] and narrowband THz spectroscopy [8] applications.

**Acknowledgement**
The support from the National Science Foundation (NSF) Award No. CBET-2006028 (S.Z. and J.-M.H.) and DMR-2237884 (Y.Z. and J.-M.H.) are gratefully acknowledged. The dynamical phase-field simulations were performed using Bridges at the Pittsburgh Supercomputing Center through allocation TG-DMR180076 from the Advanced Cyberinfrastructure Coordination Ecosystem: Services & Support (ACCESS) program, which is supported by NSF grants #2138259, #2138286, #2138307, #2137603, and #2138296. X.Z. acknowledges support from ONR YIP (N00014-23-1-2144). P. E. acknowledges support from the U. S. Department of Energy Office of Basic Energy Sciences through grant no. DE-FG02-04ER46147. Helpful discussions with Changchun Zhong and Wei Zhang are also acknowledged.

## Appendix A: Derivation of frequencies of standing acoustic phonon modes in a freestanding trilayer

The frequencies of the acoustic standing waves are obtained by solving the system of linear equations that describe the boundary conditions of the acoustic wave. Firstly, we consider an elastically heterogeneous trilayer structure as shown in Fig. 6, with $d_\xi$ being the thickness of each layer $\xi$ =A, B, C, and assume that the mechanical displacement waves propagating in the structure have only the $z$ components, considering that they are produced by the laser-induced thermal expansion. The solutions of the ($z$-component) mechanical displacement are assumed to be the combination of the plane waves $u_\xi^+ e^{i(\omega t - k_\xi z)}$ propagating along $+z$ and $u_\xi^- e^{i(\omega t + k_\xi z)}$ along $-z$ in each layer, where $u_\xi^+$ and $u_\xi^-$ are amplitudes of the wave components, $\omega$ is the angular frequency, and $k_\xi = \omega/v_\xi$ is the angular wavenumber with $v_\xi$ being the longitudinal sound speed. With these, the out-of-plane normal strain $\varepsilon_\xi(z,t) = \frac{\partial u_\xi}{\partial z}$ and stress $\sigma_\xi(z,t) = c_\xi \varepsilon_\xi(z,t)$ in each layer can be calculated as $ik_\xi[u_\xi^- e^{i(\omega t + k_\xi z)} - u_\xi^+ e^{i(\omega t - k_\xi z)}]$ and $ic_\xi k_\xi[u_\xi^- e^{i(\omega t + k_\xi z)} - u_\xi^+ e^{i(\omega t - k_\xi z)}]$, respectively, where $c_\xi$ denotes the elastic stiffness component $c_{11}$ of each layer $\xi$ here.

The stress-free boundary condition at the bottom surface of the layer A, $\sigma_A(z=0, t) = 0$, gives

$$c_A k_A (u_A^- - u_A^+) = 0. \tag{A1}$$

Similarly, the stress-free boundary condition at the top surface of the layer C, $\sigma_C(z=d, t) = 0$, gives

$$c_C k_C [u_C^- e^{ik_C d} - u_C^+ e^{-ik_C d}] = 0, \tag{A2}$$

where $d = d_A + d_B + d_C$ is the total thickness of the entire structure.

Regarding the interfaces between two elastically different materials, the stress $\sigma$ and the displacement $u$ should be continuous across the interfaces. At the A/B interface, the continuous displacement, $u_A(z=d_A, t) = u_B(z=d_A, t)$, gives

$$u_A^+ e^{-ik_A d_A} + u_A^- e^{ik_A d_A} = u_B^+ e^{-ik_B d_A} + u_B^- e^{ik_B d_A}, \tag{A3}$$

and the continuous stress, $\sigma_A(z=d_A, t) = \sigma_B(z=d_A, t)$, gives

$$c_A k_A [u_A^- e^{ik_A z} - u_A^+ e^{-ik_A z}] = c_B k_B [u_B^- e^{ik_B z} - u_B^+ e^{-ik_B z}]. \tag{A4}$$

Similarly, the continuous displacement at the B/C interface, $u_B(z=d_A+d_B, t) = u_C(z=d_A+d_B, t)$, gives

$$u_B^+ e^{-ik_B(d_A+d_B)} + u_B^- e^{ik_B(d_A+d_B)} = u_C^+ e^{-ik_C(d_A+d_B)} + u_C^- e^{ik_C(d_A+d_B)}, \tag{A5}$$

and the continuous stress, $\sigma_B(z=d_A+d_B, t) = \sigma_C(z=d_A+d_B, t)$, gives

$$c_B k_B [u_B^- e^{ik_B(d_A+d_B)} - u_B^+ e^{-ik_B(d_A+d_B)}] = c_C k_C [u_C^- e^{ik_C(d_A+d_B)} - u_C^+ e^{-ik_C(d_A+d_B)}]. \tag{A6}$$

The Equations (A1-6) above form a system of linear equations with $u_\xi^+$ and $u_\xi^-$ being six variables to be solved, and the corresponding coefficient matrix is given by,

$$\begin{pmatrix} 1 & -1 & 0 & 0 & 0 & 0 \\ e^{-ik_A d_A} & e^{ik_A d_A} & -e^{-ik_B d_A} & -e^{ik_B d_A} & 0 & 0 \\ c_A k_A e^{-ik_A d_A} & -c_A k_A e^{ik_A d_A} & -c_B k_B e^{-ik_B d_A} & c_B k_B e^{ik_B d_A} & 0 & 0 \\ 0 & 0 & e^{-ik_B(d_A+d_B)} & e^{ik_B(d_A+d_B)} & -e^{-ik_C(d_A+d_B)} & -e^{ik_C(d_A+d_B)} \\ 0 & 0 & c_B k_B e^{-ik_B(d_A+d_B)} & -c_B k_B e^{ik_B(d_A+d_B)} & -c_C k_C e^{-ik_C(d_A+d_B)} & c_C k_C e^{ik_C(d_A+d_B)} \\ 0 & 0 & 0 & 0 & e^{-ik_C d} & -e^{ik_C d} \end{pmatrix}.$$

To obtain non-trivial solution to linear equations system, the determinant of this matrix should be equal to 0, which gives the following equation,

$$c_A c_C \left(-1+e^{\frac{2id_A\omega}{v_A}}\right)\left(-1+e^{\frac{2id_B\omega}{v_B}}\right)\left(-1+e^{\frac{2id_C\omega}{v_C}}\right) v_B^2$$

$$+c_B^2 \left(1+e^{\frac{2id_A\omega}{v_A}}\right)\left(-1+e^{\frac{2id_B\omega}{v_B}}\right)\left(1+e^{\frac{2id_C\omega}{v_C}}\right) v_A v_C$$

$$+c_B c_C \left(1+e^{\frac{2id_A\omega}{v_A}}\right)\left(1+e^{\frac{2id_B\omega}{v_B}}\right)\left(-1+e^{\frac{2id_C\omega}{v_C}}\right) v_A v_B$$

$$+c_A c_B \left(-1+e^{\frac{2id_A\omega}{v_A}}\right)\left(1+e^{\frac{2id_B\omega}{v_B}}\right)\left(1+e^{\frac{2id_C\omega}{v_C}}\right) v_B v_C = 0, \qquad (A7)$$

where all wavenumbers $k_\xi$ are replaced by $\omega/v_\xi$.

# Appendix B. Expressions of Ω and Λ in the analytical formula of the dispersion relation of exchange magnon

Assume the magnetization at initial equilibrium state is $(m_x^0, m_y^0, m_z^0)$, the expressions of the $\Omega$ and $\Lambda$ are given by,

$$\Omega = (\Psi_{23}-\Psi_{32})m_x^0+(\Psi_{31}-\Psi_{13})m_y^0+(\Psi_{12}-\Psi_{21})m_z^0, \tag{B1}$$

and

$$\Lambda = \Psi_{23}\Psi_{32}+\Psi_{31}\Psi_{13}+\Psi_{12}\Psi_{21}, \tag{B2}$$

where

$$\Psi_{12}=H_z^{\text{bias}}-M_s m_z^0 - \frac{2K_1}{\mu_0 M_s}\left(3m_y^{0^2}m_z^0 - m_z^{0^3}\right), \tag{B3}$$

$$\Psi_{13}=-H_y^{\text{bias}}-M_s m_y^0 - \frac{2K_1}{\mu_0 M_s}\left(m_y^{0^3} - 3m_y^0 m_z^{0^2}\right), \tag{B4}$$

$$\Psi_{21}=-H_z^{\text{bias}}+M_s m_z^0 - \frac{2K_1}{\mu_0 M_s}\left(m_z^{0^3} - 3m_z^0 m_x^{0^2}\right), \tag{B5}$$

$$\Psi_{23}=H_x^{\text{bias}}+M_s m_x^0 - \frac{2K_1}{\mu_0 M_s}\left(3m_z^{0^2}m_x^0 - m_x^{0^3}\right), \tag{B6}$$

$$\Psi_{31}=H_y^{\text{bias}} - \frac{2K_1}{\mu_0 M_s}\left(3m_x^{0^2}m_y^0 - m_y^{0^3}\right), \tag{B7}$$

$$\Psi_{32}=-H_x^{\text{bias}} - \frac{2K_1}{\mu_0 M_s}\left(m_x^{0^3} - 3m_x^0 m_y^{0^2}\right), \tag{B8}$$

with $K_1$ being magnetocrystalline anisotropy coefficient. The magnon dispersion relation $\omega_m(k_m)$ is obtained analytically from the linearization of the LLG equation under zero magnetic damping ($\alpha=0$). Detailed procedures can be found in our previous work [15]

## Appendix C. Acoustic excitation of magnons in 2D multiphase system

The laser-induced acoustic pulse injection is assumed to be uniform on the Pt top surface. As shown in Fig. 7(a), the strain $\varepsilon_{zz}$ is almost uniform along the $x$ axis after the standing acoustic wave forms inside the Pt/(001)MAFO/SiN. As a result, the acoustically excited magnons $\Delta m_x(\mathbf{r}, t)$ ($\mathbf{r}=x,y,z$) is also almost uniform along the $x$ axis, as shown in Fig. 7(c). At near the stress-free lateral surfaces of the Pt and the MAFO layers, acoustic wave with mechanical displacement $u_x$ is produced and propagates along the $x$ axis due to the time-varying tension (positive $\varepsilon_{zz}$) and compression (negative $\varepsilon_{zz}$) along the $z$ axis from the strain injection. This induces a non-uniform in-plane strain $\varepsilon_{xx}$ and hence the $\varepsilon_{zz}$ in the region which the acoustic wave $u_x$ propagated across, as shown in Fig. 7(b). Accordingly, the distribution of magnons $\Delta m_x(\mathbf{r}, t)$ is also nonuniform near the lateral surfaces of Pt and MAFO, as shown in Fig. 7(d). However, the standing acoustic waves and the excited magnons in the most regions of the membrane are uniform along the $x$ axis. This is because, in MAFO with an in-plane size of 1×1 mm$^2$ for example, the $\varepsilon_{xx}$ can only travel for ~20 μm in 2 ns from the lateral surfaces of MAFO towards its center, even if we use a high longitudinal sound speed of 9462 m s$^{-1}$ (the same as that in SiN). This means that the phonons and the magnons in 92% [=(1 mm - 0.02 mm×2)$^2$/1 mm$^2$] of the area should still be uniform in the $xy$ plane in 2 ns.

The simulation results above were obtained by discretizing the entire heterostructure (including the freestanding multilayer, the air, and the Si substrate) into a 2D system of computational cells. The numbers of the computational cells along $x$ and $z$ axes are $n_x$ = 50,000 and $n_z$ = 93, respectively. The cell size along $x$ and $z$ axes are $\Delta x$ = 1 nm and $\Delta z$ = 0.83 nm, respectively. Starting from the top surface of the simulation system, the top 2 layers of the cells (2$\Delta z$) are designated as the free space. The following Pt and MAFO layers are discretized into 8 layers and 15 layers of the computational cells along $z$ axis, respectively. Next, the following 38 layers of the computational cells are all designated as SiN. In the last 30 layers of the computational cells, the bottom 20 layers are all designated as the Si substrate and the remaining 10 layers are designated as either Si or air as shown in Fig. 7(a), which makes the SiN bottom surface stress-free and enables it to reflect the acoustic wave. The air below the SiN, Pt and MAFO layers are all discretized into 40,000 computational cells along the $x$ axis.

In the 2D simulations above, the dynamics of the EM waves (Eqs. (4-5)) and hence the induced eddy current (Eq. (7)) are not considered. The LLG equation (Eq. (2) in the main text) and the elastodynamic equation (Eq. (4)) are solved in a coupled fashion with a real-time step $\Delta t$ = 2×10$^{-15}$ s. For the boundary conditions of the elastodynamics in the 2D system, $\sigma_{ix}$ = 0 ($i=x,y,z$) is applied on the stress-free lateral surfaces of the Pt and MAFO layers, and $\sigma_{iz}$ = 0 is applied on the stress-free top surface of the Pt layer. To model the injection of ps-duration acoustic pulse, a Gaussian-shaped stress pulse $\sigma_{zz}(z = d_{\text{MAFO}}+d_{\text{Pt}}, t)$ = $\sigma_{\text{max}}\exp[-\frac{t^2}{2\tau^2}]$ is applied at the top surface of the Pt layer at $t$=0 ps, where $\sigma_{\text{max}}$=3 GPa and $\tau$=1.5 ps are the same as those used in Fig. 3. The absorbing boundary condition, $\frac{\partial u_i}{\partial z}=-\frac{1}{v}\frac{\partial u_i}{\partial t}$ ($i=x,y,z$), is applied at the bottom surface of the Si substrate to make it a perfect sink for acoustic waves. Here $v$ is the transverse sound velocity in Si for $u_x$ and $u_y$ and the longitudinal sound velocity for $u_z$. The magnetic boundary condition $\partial \mathbf{m}/\partial \mathbf{n}$ = 0 [63] is applied on all surfaces of the MAFO layer where $\mathbf{n}$ is unit vector normal to the surface. The physical validity of our in-house 2D elastodynamic solver is demonstrated by benchmarking test against the results obtained from COMSOL Multiphysics, as shown in Fig. 8.

# Appendix D. Derivation of the interfacial spin current density and the resulting iSHE charge current density

Equation (8) can be expanded into,

$$\mathbf{e}_n \cdot \mathbf{J}^{s,0} = [0,0,+1] \cdot \begin{bmatrix} J_{xx}^{s,0} & J_{xy}^{s,0} & J_{xz}^{s,0} \\ J_{yx}^{s,0} & J_{yy}^{s,0} & J_{yz}^{s,0} \\ J_{zx}^{s,0} & J_{zy}^{s,0} & J_{zz}^{s,0} \end{bmatrix} = [J_{zx}^{s,0}, J_{zy}^{s,0}, J_{zz}^{s,0}]$$

$$= \frac{\hbar}{4\pi} G_{\text{eff}}^{\uparrow\downarrow} \left[ \left( m_y \frac{\partial m_z}{\partial t} - m_z \frac{\partial m_y}{\partial t} \right), \left( m_z \frac{\partial m_x}{\partial t} - m_x \frac{\partial m_z}{\partial t} \right), \left( m_x \frac{\partial m_y}{\partial t} - m_y \frac{\partial m_x}{\partial t} \right) \right]. \quad (D1)$$

Since $J_x^{\text{iSHE}}(z,t) = -\theta_{\text{Pt}} \frac{2e}{\hbar} J_{zy}^s(z,t)$, $J_y^{\text{iSHE}}(z,t) = \theta_{\text{Pt}} \frac{2e}{\hbar} J_{zx}^s(z,t)$, and $J_z^{\text{iSHE}}(z,t) = 0$, one can further write,

$$\begin{bmatrix} J_x^{\text{iSHE},0} \\ J_y^{\text{iSHE},0} \\ J_z^{\text{iSHE},0} \end{bmatrix} = \begin{bmatrix} \frac{\theta_{\text{Pt}} G_{\text{eff}}^{\uparrow\downarrow} e}{2\pi} \left( m_x \frac{\partial m_z}{\partial t} - m_z \frac{\partial m_x}{\partial t} \right) \\ \frac{\theta_{\text{Pt}} G_{\text{eff}}^{\uparrow\downarrow} e}{2\pi} \left( m_y \frac{\partial m_z}{\partial t} - m_z \frac{\partial m_y}{\partial t} \right) \\ 0 \end{bmatrix}, \quad (D2)$$

where $\mathbf{m}$ and $\frac{\partial \mathbf{m}}{\partial t}$ are based on the time-varying magnetization at the MAFO/Pt interface. If omitting the damping, one can write under plane-wave assumption that,

$$m_i = m_i^{\text{eq}} + \Delta m_i; \quad \Delta m_i = |\Delta m_i^0| e^{i(kz - \omega t)}, i = x, y, z, \quad (D3)$$

for a specific magnon mode, or $\Delta m_i(z,t) = |\Delta m_i^0| \cos(kz - \omega t)$ in real space and time domain, where $k$ is the wavenumber and $|\Delta m_i^0|$ is the peak amplitude of magnetization component variation. $m_i^{\text{eq}}$ are the normalized magnetization component at the initial equilibrium, with $(m_x^{\text{eq}}, m_y^{\text{eq}}, m_z^{\text{eq}}) = (\frac{1}{2}, \frac{1}{2}, \frac{\sqrt{2}}{2})$ herein. Plugging Eq. (D3) into Eq. (D2), the $J_i^{\text{iSHE},0}$ can be further written as,

$$\begin{bmatrix} J_x^{\text{iSHE},0} \\ J_y^{\text{iSHE},0} \\ J_z^{\text{iSHE},0} \end{bmatrix} = \begin{bmatrix} \frac{\theta_{\text{Pt}} G_{\text{eff}}^{\uparrow\downarrow} e}{2\pi} (m_x^{\text{eq}} |\Delta m_z^0| - m_z^{\text{eq}} |\Delta m_x^0|) \omega \sin(kz - \omega t) \\ \frac{\theta_{\text{Pt}} G_{\text{eff}}^{\uparrow\downarrow} e}{2\pi} (m_y^{\text{eq}} |\Delta m_z^0| - m_z^{\text{eq}} |\Delta m_y^0|) \omega \sin(kz - \omega t) \\ 0 \end{bmatrix}. \quad (D4)$$

Thus, the peak magnitude of the $J_i^{\text{iSHE},0}$ depends not only on the $|\Delta m_i^0|$ but also the angular frequency $\omega$. Based on the simulated $\Delta m_i(z = d_{\text{MAFO}}, t)$, we extract $(|\Delta m_x^0|, |\Delta m_y^0|, |\Delta m_z^0|) = (0.085, 0.0849, 0.0693)$ for the 154 GHz $m=1$ mode magnon (corresponding to Fig. 3(c)) and $(|\Delta m_x^0|, |\Delta m_y^0|, |\Delta m_z^0|) = (0.0672, 0.0669, 0.0546)$ for the 240 GHz $m=1$ magnon mode (corresponding to Fig. 4(d)). Based on Eq. (D4), one can predict, for example, the peak magnitude of $J_x^{\text{iSHE},0}$ resulting from the 240 GHz m=1 magnon mode is about 1.24 times larger than that from the 154 GHz m=1 magnon mode. This is consistent with the ratio of 1.22 obtained from the time-domain profile of the filtered $J_x^{\text{iSHE},0}$ data, as shown by Fig. 9. It is noteworthy that the 240-GHz iSHE charge current density has a larger peak amplitude than the 154-GHz counterpart but a faster attenuation.

**Appendix E. A list of material parameters used in analytical calculation and dynamical phase-field simulation**

The following parameters for (001) MAFO are taken from [29,64], including the Gilbert damping coefficient $\alpha^0$=0.0015 for m=0 magnon mode, mass density $\rho$ = 4355 kg m$^{-3}$; gyromagnetic ratio $\gamma$ = 0.227 rad MHz A$^{-1}$ m; saturation magnetization $M_s$ = 0.0955 MA m$^{-1}$; magnetocrystalline anisotropy coefficient $K_1$ = -477.5 J m$^{-3}$; and magnetoelastic coupling coefficients $B_1$ = 1.2 MJ m$^{-3}$ and $B_2$=0 [29]. The elastic stiffness coefficients of the MAFO, $c_{11}$ = 282.9 GPa, $c_{12}$ = 155.4 GPa, and $c_{44}$ = 154.8 GPa, are assumed to be the same as those of MgAl$_2$O$_4$ [65], while the exchange coupling coefficient $A_{ex}$=4 pJ m$^{-1}$ is assumed to be same as that of CoFe$_2$O$_4$ [39].

For SiN: $c_{11}$ = 283.81 GPa, $c_{12}$ = 110.37 GPa and $c_{44}$ = 86.72 GPa are calculated using its Young's modulus of 222 GPa and Poisson's ratio of 0.28 [66] by assuming isotropic elasticity, as is appropriate for an amorphous solid. The mass density $\rho$ = 3170 kg m$^{-3}$.

For Si [67] which was incorporated as the supporting substrate in the 2D simulations (see Appendix C): $c_{11}$ = 167.4 GPa, $c_{12}$ = 65.2 GPa, $c_{44}$ = 79.6 GPa and $\rho$ = 2330 kg m$^{-3}$.

For Pt [68]: $c_{11}$ = 347 GPa, $c_{12}$ = 250 GPa, $c_{44}$ = 75 GPa and $\rho$ = 21450 kg m$^{-3}$. The plasma frequency $\omega_p$ = 9.1 rad fs$^{-1}$ and electron relaxation time $\tau_e$ = 7.5 fs are taken from [69].

Acoustic attenuation data for MAFO is not yet available so the phenomenological stiffness damping coefficients $\beta$ of all materials are assumed to be same as that of the Si substrate, with $\beta$ = 4.48×10$^{-15}$ s which is comparable to the value previously extracted for (001) Ga$_3$Gd$_5$O$_{12}$ single crystals [16]. The value of $\beta$ for Si was obtained by fitting the experimentally determined attenuation coefficient $\lambda$ = 9 cm$^{-1}$ of a 7.2 GHz transverse acoustic wave in Si [67] to an analytical formula, $\beta = \frac{2k\lambda}{\omega(k^2-\lambda^2)}$ [16], where the $\omega$ = 2$\pi$×7.2 GHz and the $k = \omega/v$ are the angular frequency and angular wavenumber, respectively; and $v$=5090 m/s is the longitudinal speed of sound in Si [67]. From the formula for $\beta(k)$, we calculate the $\lambda$ for other frequencies and wavenumbers, which enables the analytical calculation of the lifetime of the acoustic phonon modes $\tau_{ph}$=1/($\lambda v$). The relative permittivity $\varepsilon_r$ is assumed to be 1 for all materials.

**Appendix F. Numerical methods for solving the coupled LLG, elastodynamic, and Maxwell's equations**

For the simulations of Figs. 3-5, the system is discretized into a 1D grid of computational cells along the $z$ axis, with a cell size $\Delta z = 0.83$ nm (in Figs. 3 and 4) or $\Delta z = 0.3$ nm (in Fig. 5). Equations (2-6) are solved simultaneously with a real-time step $\Delta t = 2\times10^{-18}$ s. When solving the equations, the central finite difference method is used to numerically calculate the spatial derivatives and the classical Runge-Kutta method is used for time marching.

When solving the LLG equation (Eq. (2)), the magnetic boundary condition $\partial \mathbf{m}/\partial \mathbf{n} = 0$ [63] is applied on all surfaces of the MAFO where $\mathbf{n}$ is unit vector normal to the surface. When solving the elastodynamic equation (Eq. (4)), the continuity of the mechanical displacement $\mathbf{u}$ and stress $\boldsymbol{\sigma}$ are applied at any interface between two elastically different materials. The boundary condition of continuous stress at Pt top surface and SiN bottom surface becomes stress-free boundary condition since the stress $\boldsymbol{\sigma}$ in the free space is 0, specifically, $\sigma_{iz} = 0$ ($i=x,y,z$). As mentioned in the main text, the injection of the $ps$ bulk acoustic pulse $\varepsilon_{zz}(z,t)$ is simulated by applying a time-varying stress $\sigma_{zz}(t)$ (time-dependent boundary condition) at the top surface of the Pt layer. Note that the applied stress $\sigma_{zz}(t)$ converges to 0 in the course of time, which enables the top surface of the Pt layer to be stress-free again after the injection of the $ps$ acoustic pulse. For substrate-supported Pt/MAFO/SiN multilayer (the control simulation), the absorbing boundary condition, $\frac{\partial u_i}{\partial z}=-\frac{1}{v}\frac{\partial u_i}{\partial t}$ ($i=x,y,z$), is applied at the top surface of the SiN layer to make it a perfect sink for acoustic waves. Here $v$ is the transverse sound velocity for $u_x$ and $u_y$ and the longitudinal sound velocity for $u_z$.

Maxwell's equations (Eqs. (5-6)) are solved using the conventional finite-difference time-domain (FDTD) method. In the 1D system, the absorbing boundary condition $\frac{\partial \mathbf{E}^{EM}}{\partial z}=-\frac{1}{c}\frac{\partial \mathbf{E}^{EM}}{\partial t}$ [70] is applied on both the bottom and top surfaces of the computational system to prevent the emitted EM waves from being reflected to the system, where $c$ is the light speed in the free space.

### Appendix G. Temporal profiles of the *m*=1 mode magnon extracted from Fig. 3(c)

Using inverse Fourier transform (similarly to the procedures in Fig. 9), we extract the temporal profiles of the *m*=1 mode magnon in both the freestanding and substrate-supported multilayer, as shown in Fig. 10.

### Appendix H. Temporal profiles of $\mathbf{J}^{iSHE}(t)$ and $\mathbf{J}^{p}(t)$ at the Pt/MAFO interface corresponding to the $\mathbf{J}^{c}(t)$ in Fig. 3(e)

The polarization (eddy) current $\mathbf{J}^{p}(t)$ has a 180° phase difference with the $\mathbf{J}^{iSHE}(t)$ and not negligible, as shown in Fig. 11. As a result, the amplitude of $\mathbf{J}^{c}(t) = \mathbf{J}^{iSHE}(t) + \mathbf{J}^{p}(t)$ is smaller than that of the $\mathbf{J}^{iSHE}(t)$.

### Appendix I. Temporal profile and frequency spectrum of the free-space $\mathbf{E}^{EM}(t)$ corresponding to the $\mathbf{J}^{c}(t)$ in Fig. 3(e)

Both the temporal profile and the frequency spectrum of the emitted electric field $\mathbf{E}^{EM}(t)$ are similar to those of $\mathbf{J}^{c}(t)$, as shown in Fig. 12.

### Appendix J. Nonlinear magnon modes induced by the acoustic phonon modes in Fig. 4(c)

Under multiple driving standing acoustic phonon modes (up to the sixth order) in the freestanding multilayer, the acoustically excited magnon modes display complex frequency mixing and doubling behaviors, as shown in Fig. 13.

**Appendix K. Tabular data related to Fig. 5 and extended discussion**

In the substrate-supported multilayer, the absorbing boundary condition is applied to the bottom surface of the SiN layer to make the SiN a perfect acoustic sink. The dominant magnon mode in the freestanding multilayer has the same frequency as the driving standing acoustic phonon mode ($f_n$), while the dominant magnon mode in the substrate-supported multilayer has the same frequency as the $m=1$ mode exchange magnon ($f_{m=1}$). Under the exact resonant magnon-phonon condition, one has $f_n = f_{m=1}$. As shown in Table 1, in some cases, there exists a small discrepancy between $f_n$ and $f_{m=1}$ because (i) the $d_{MAFO}$ and $d_{SiN}$ used in numerical simulations cannot exactly match those predicted analytically via Eq. (1); and (ii) the unavoidable numerical error in discrete Fourier transform. Here, both the frequency $f$ and linewidth $\Delta f$ are extracted from the frequency spectra of the time-domain iSHE charge current density $J_x^{iSHE}(t)$ data, rather than the total $J_x^c(t) = J_x^{iSHE}(t) + J_x^p(t)$ at the MAFO/Pt interface. Our tests have shown that the $Q$ extracted from the frequency spectra of $J_x^{iSHE}(t)$ and $J_x^c(t)$ are almost the same. The $J_x^{iSHE}(t)$ can be acquired from coupled phonon-magnon dynamics simulations, while obtaining the $J_x^c(t)$ requires the use of coupled phonon-magnon-photon dynamics simulations that are computationally more expensive.

# Appendix L. Influence of elastic damping and magnetic damping on the quality factor of the freestanding multilayer

As shown in Fig. 14(a), turning off either type of damping enhances both the amplitude and the lifetime of the $J_x^c$. A longer lifetime results in an increase in the quality factor ($Q$). Moreover, turning off the magnetic damping (gray curve) leads to a larger peak amplitude for the $J_x^c$ than the case of turning off elastic damping (green curve). This is reasonable because $J_x^c$ results from the spin current density and hence the magnetization amplitude at the MAFO/Pt interface. However, turning off the elastic damping (green curve) leads to an even longer lifetime, which should lead to a larger $Q$. This expectation is consistent with the frequency spectra in Fig. 14(b), which shows that the green curve (zero elastic damping) has the narrowest linewidth and thus the largest $Q$.

In our numerical simulations, the layer thickness must be equal to $N\Delta z$, where $\Delta z=0.83$ nm is the cell size, and the integer number $N$ is the number of cells. Under the present thickness set up (see Fig. 14 caption), the frequencies of the $n=2$ mode acoustic phonon and $m=1$ mode exchange magnon are 153.8 GHz and 154.407 GHz, respectively, which are calculated analytically via Eq. (1) in the main paper. As shown in Fig. 14(b), when the magnetic damping is ON (green and blue curves), the $m=1$ mode exchange magnon will decay and vanish. The dominant magnon mode in the freestanding multilayer is the 'acoustic' magnon mode, which is excited by acoustic pulse via the magnetoelastic coupling and therefore has the same frequency as the driving $n=2$ mode acoustic phonon (the simulated peak at 153.75 GHz agrees very well with the analytical calculation). When the magnetic damping is OFF (gray curve), the $m=1$ mode exchange magnon, with a peak at 154.4 GHz (which is again almost the same as analytically calculated value), is significantly amplified and becomes the dominant magnon mode instead.

**Appendix M. Modeling the formation of THz magnon polarons**

The thicknesses of the Pt, MAFO and SiN films are set as 6.6 nm, 5.9 nm and 18.1 nm, respectively, to enable the formation of $n$=6 mode phonon and $m$=1 mode magnon, which have the same frequency of 682.8 GHz. Meanwhile, the phonon and the magnon have almost the same spatial profile in the MAFO film (as shown in Fig. 15(a)), leading to the same wavenumber of the two waves and the formation of magnon polaron. The formation of the magnon polaron is demonstrated by the two frequency peaks (678 GHz and 688 GHz) split from the 682.8 GHz in the spectra of both the magnon and phonon (as shown in Fig. 15(e)). In this simulation, the free parameters in the $\sigma_{zz}(t)$ (see Sect. IV) are $\sigma_{max}$=3 MPa and $\tau$ =0.5 ps, which lead to a frequency window covering up to 800 GHz and a peak amplitude of $8.5\times10^{-6}$ in the strain pulse $\varepsilon_{zz}(t)$ in the Pt. The effective magnetic damping $\alpha$ and the stiffness damping coefficients $\beta$ are both set to be 0, and a scaled-up magnetoelastic coupling coefficient $B_1$=25$B_1^0$ is used to enhance the magnon-phonon coupling strength, where $B_1^0$=1.2 MJ m$^{-3}$ is the value given in Appendix E.

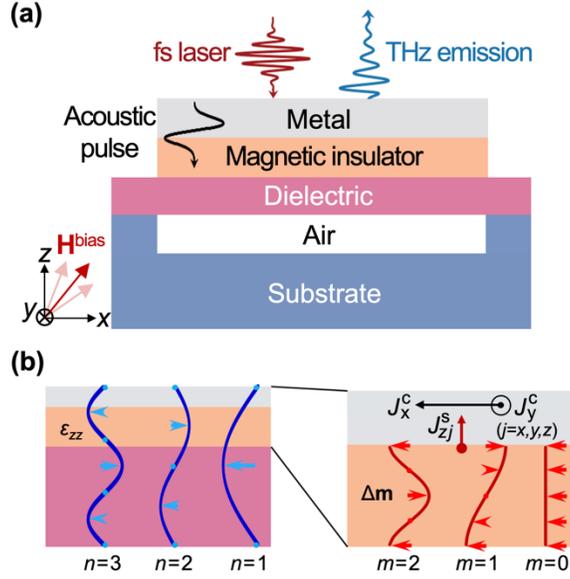

**Figure 1.** **(a)** Schematic of the freestanding multilayer on a substrate patterned with air cavity at its surface. The acoustic pulse is injected into the multilayer by irradiating the metal film with a femtosecond (fs) laser pulse. **(b)** Illustration of the standing longitudinal acoustic phonons $\varepsilon_{zz}$ with modes of $n$ in the multilayer, and the standing magnons $\Delta\mathbf{m}$ with modes of $m$ in the magnetic insulator film. $J^s_{zj}$ is the spin current tensor injected from the magnet/metal interface, where the subscript '$z$' and '$j$'($=x,y,z$) refer to the direction of spin current flow and spin polarization, respectively. $J^c_i$ is the charge current vector in the metal layer converted from $J^s_{zj}$ via the inverse Spin Hall effect. A bias magnetic field $\mathbf{H}^{bias}$ with nonzero $z$-component is applied to (i) lift the initial equilibrium magnetization off the $xy$ plane (for enhancing the torque from the magnetoelastic field) and (ii) dynamically tune the frequencies of the ferromagnetic resonance ($m$=0 mode magnon) and thereby higher-order magnons. The acoustically excited magnons can be detected by measuring the $J^c_i$ or the resulting THz emission.

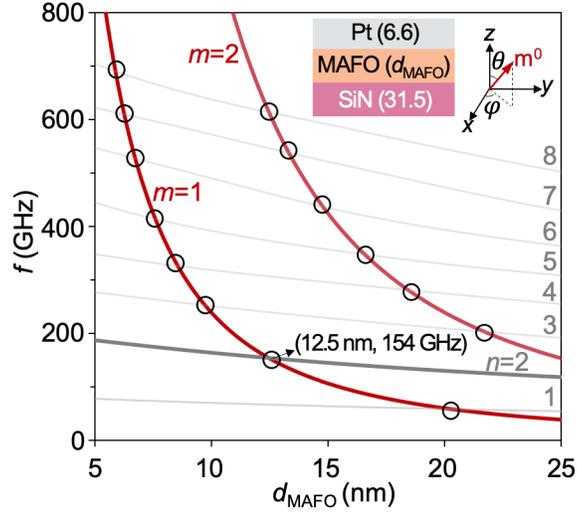

**Figure 2**. Analytically calculated frequencies of the standing $m=1$ and $m=2$ magnon modes and the $n=1,2…8$ acoustic phonon modes, as a function of the magnetic insulator (MAFO) film thickness $d_{MAFO}$. The thicknesses for the Pt and SiN layers are fixed and indicated by the numbers in parentheses with a unit of nm. Another inset indicates the tilted initial equilibrium magnetization m0, with $\theta=\varphi=45°$ in this calculation.

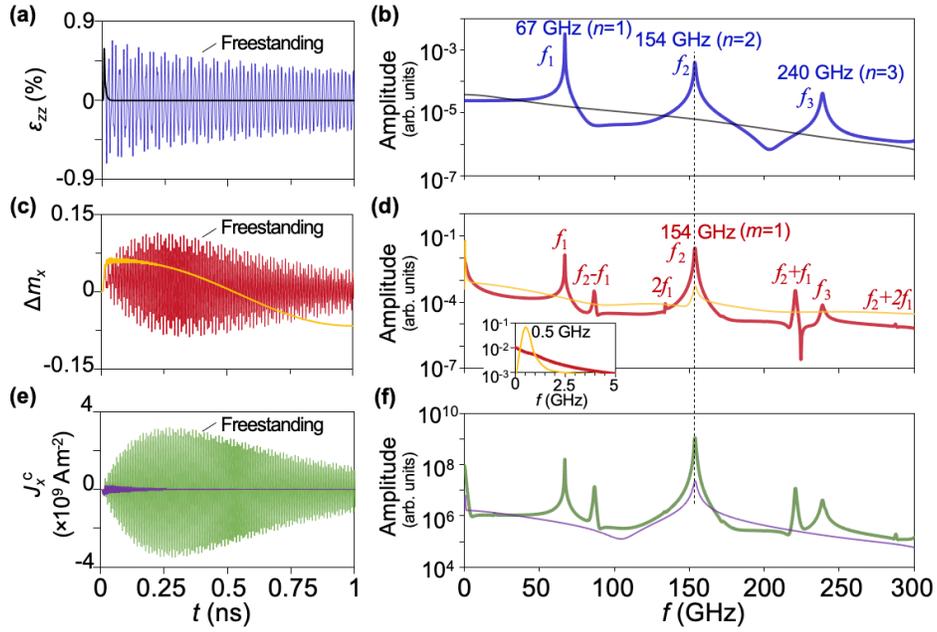

**Figure 3**. Evolution of **(a)** the strain $\varepsilon_{zz}(t)$, **(c)** the normalized magnetization change $\Delta m_x(t)=m_x(t)-m_x(t=0)$, and **(e)** the charge current density $J_x^c(t)$ at the Pt/MAFO interface in both a freestanding Pt/MAFO/SiN and a substrate-supported Pt/MAFO/SiN(acoustic sink) multilayer. $t=0$ is the moment the acoustic pulse was injected from the top surface of the Pt film. **(b),(d),(f)** Frequency spectra of the $\varepsilon_{zz}(t)$, $\Delta m_x(t)$, and $J_x^c(t)$, respectively. The vertical axes of these sub-figures are plotted in log scale. The vertical dashed line emphasizes the resonant interaction between $n=2$ phonon mode and $m=1$ magnon mode at 154 GHz.

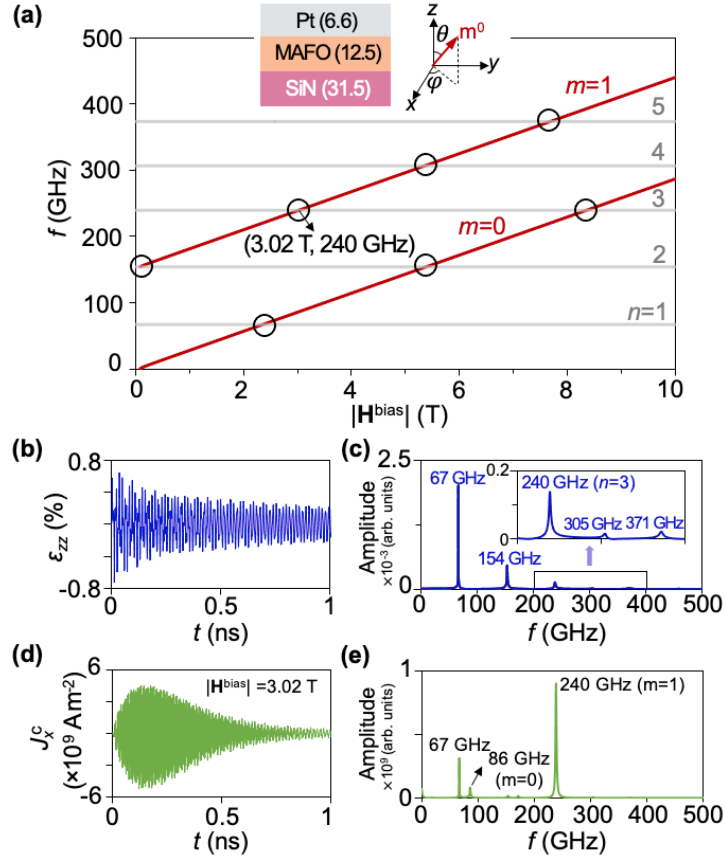

**Figure 4**. **(a)** Analytically calculated frequencies of the *m*=0 and *m*=1 magnon modes as functions of the magnitude of bias magnetic fields $\mathbf{H}^{\text{bias}}$, with the frequencies of the standing acoustic phonon modes *n*=1 to 5. The circles indicate the $|\mathbf{H}^{\text{bias}}|$ that leads to resonant magnon-phonon interaction. Evolution of **(b)** the strain $\varepsilon_{zz}(t)$ and **(d)** the charge current density $J_x^c(t)$ at the Pt/MAFO interface. **(c),(e)** Their corresponding frequency spectra where the vertical axes are plotted in linear scale.

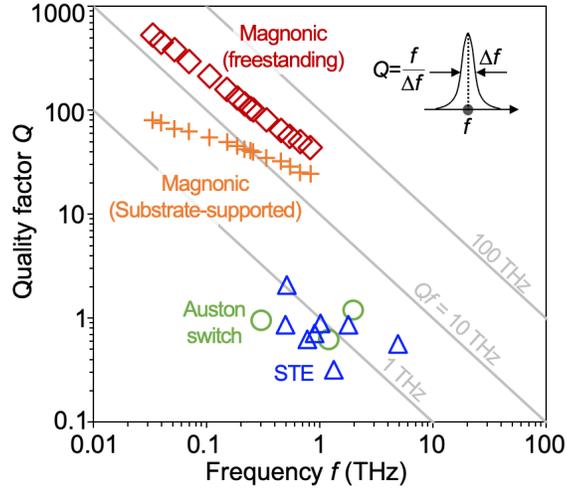

**Figure 5.** Simulated quality factor $Q$ vs. resonant frequency $f$ of the magnonic optoelectronic transducer based on both the freestanding and substrate-supported multilayer. Typical $Q$ values of the Auston switch and the spintronic THz emitter (STE), both of which permits the conversion of a fs optical pulse into a THz charge current pulse as well. The inset illustrates the definition of $Q$. Literature sources of the data points for Auston switch and STE are provided in [8].

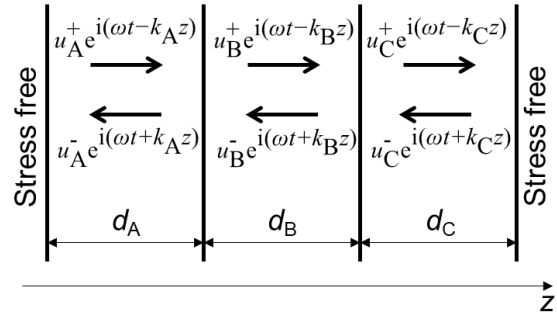

**Figure 6.** Schematic of an elastically heterogeneous trilayer membrane structure where mechanical displacement waves **u**(*z*,*t*) are propagating.

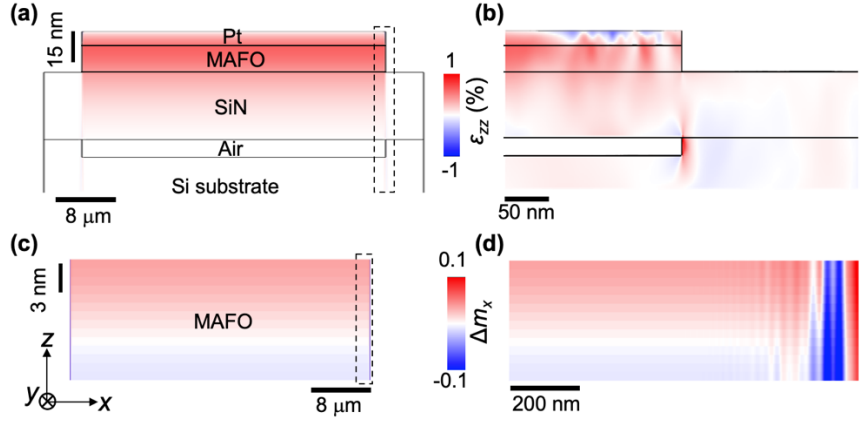

**Figure 7**. (a) Distribution of strain $\varepsilon_{zz}$ in $xz$ cross section of the freestanding Pt/(001) MAFO/SiN multilayer at $t$ = 54 ps after the strain injection, and (b) enlarged section of the dashed rectangles in (a). The freestanding multilayer is integrated on Si substrate with air cavity at its surface. (c) Distribution of magnetization change $\Delta m_x$ in $xz$ cross section of the MAFO layer at $t$ = 54 ps after the laser pulse excitation, and (d) enlarged section of the dashed rectangles in (c). All physical quantities are assumed to be uniform along the $y$ axis.

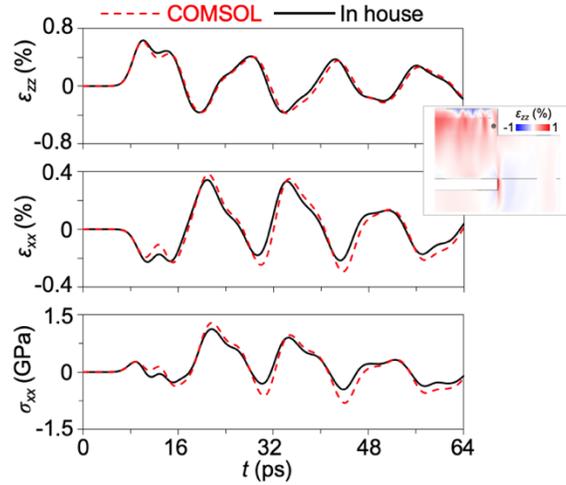

**Figure 8.** Comparison between the simulations from COMSOL Multiphysics and our in-house solver in temporal evolution of the out-of-plane strain $\varepsilon_{zz}$, in-plane strain $\varepsilon_{xx}$, and the in-plane stress $\sigma_{xx}$ at the cell illustrated by the dot (in the middle layer of the MAFO film, 10 nm away from the edge) in the inset strain distribution (which is a section of Fig. 7(b)).

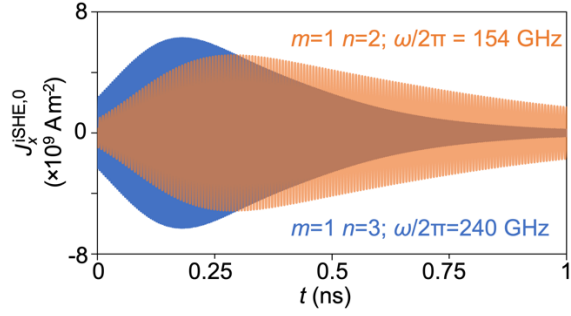

**Figure 9**. Evolution of the filtered $J_x^{iSHE,0}$ resulting from the 154 GHz $m=1$ mode magnon (orange, contributing to the total $J_x^c$ in Fig. 3(e)) and the 240 GHz $m=1$ magnon mode (blue, corresponding to the total $J_x^c$ in Fig. 4(d)). In the filtering process, a third-order bandpass filter, with a bandwidth of 5 GHz, was employed to isolate the frequency components near 154 GHz (i.e., from 151.5 GHz to 156.5 GHz) and near 240 GHz (i.e., from 237.5 GHz to 242.5 GHz) from time-domain data. These two frequency ranges were carefully chosen to capture the target signal while minimizing the extraneous frequency interference. A zero-phase technique was also applied to maintain the temporal integrity of the original signal.

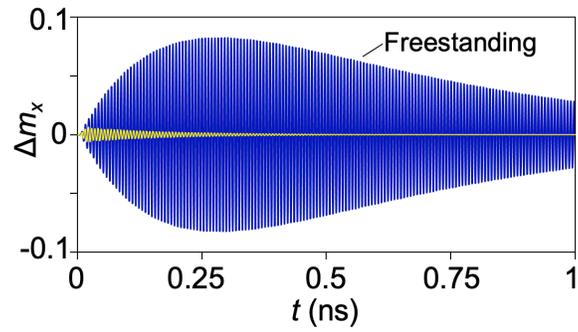

**Figure 10.** Evolution of the extracted magnetization variation $\Delta m_x(t)$ for the $m=1$ magnon mode in the freestanding Pt/MAFO/SiN multilayer (blue) and substrate-supported (yellow) Pt/MAFO/SiN(substrate, acoustic sink) heterostructure.

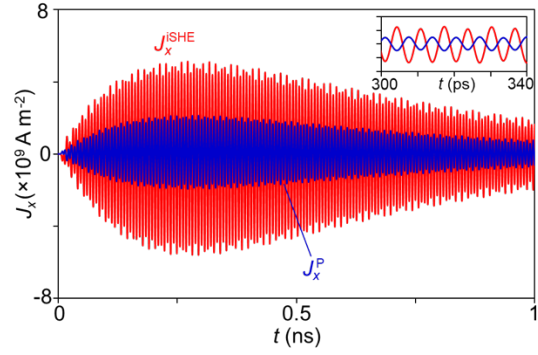

**Figure 11.** Evolution of the iSHE charge current $\mathbf{J}^{iSHE}$ (in red) and the polarization current $\mathbf{J}^{p}$ (in blue) at the Pt/MAFO interface ($z = d_{MAFO}$) in the freestanding multilayer. The inset enlarges their evolutions during $t = 300\text{-}340$ ps to show the phase difference between them.

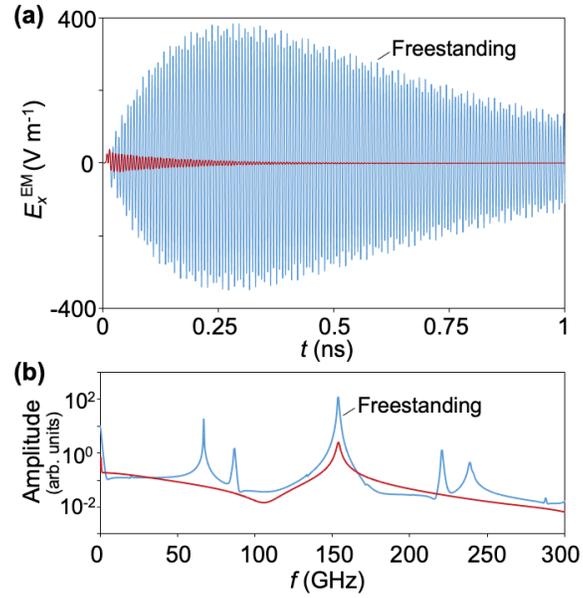

**Figure 12.** (a) Evolution of electric field component $\mathbf{E}^{EM}(t)$ of the EM wave in the free space, or more specifically, at 4 nm above the Pt top surface ($z=d_{MAFO}+d_{Pt}+4$ nm), which are emitted from the freestanding (blue) and substrate-supported (red) Pt/MAFO/SiN multilayer. (b) Frequency spectra of the $\mathbf{E}^{EM}(t)$ data.

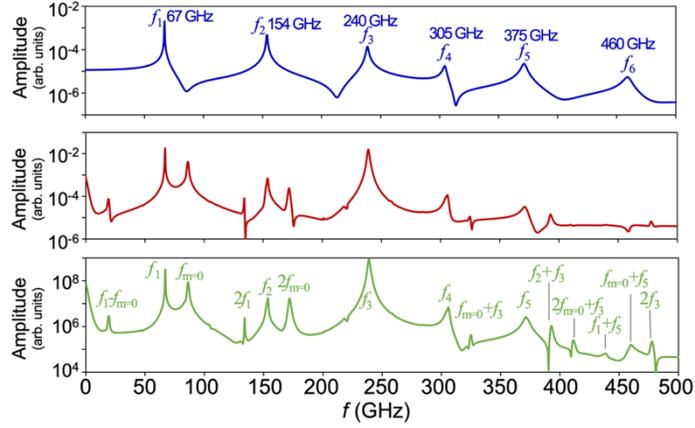

**Figure 13.** Frequency spectra of (top) strain $\varepsilon_{zz}(t)$, (middle) magnon component $\Delta m_x(t)$, and (bottom) total charge current density $J_x^c(t)$ at the Pt/MAFO interface, where the vertical axes are plotted in log scale. The time-domain data of $\varepsilon_{zz}(t)$ and $J_x^c(t)$ are shown in Figs. 4(b) and 4(d).

**Table 1**. Thicknesses of the MAFO and SiN layer used in Fig. 5, the simulated resonant frequency $f$, linewidth $\Delta f$, and $Q$ ($=f/\Delta f$) for both the freestanding and substrate-supported multilayer. The thickness of Pt is fixe at 6.6 nm.

| $d_{MAFO}$ (nm) | $d_{SiN}$ (nm) | $\alpha$ (magnetic damping) | $f=f_n$ (GHz) | $\Delta f$ (GHz) | $Q$ | $f=f_{m=1}$ (GHz) | $\Delta f$ (GHz) | $Q$ |
|---|---|---|---|---|---|---|---|---|
| | | | \multicolumn{3}{c}{Freestanding multilayer} | \multicolumn{3}{c}{Substrate-supported multilayer} | | |
| 5.4 | 7.8 | 0.01117797 | 814.5 | 18.6152421 | 43.75446727 | 814.75 | 33.1032125 | 24.6124149 |
| 6 | 19.2 | 0.01037147 | 659.5 | 13.1166238 | 50.27970689 | 659.5 | 25.6598802 | 25.7016009 |
| 6.6 | 12.6 | 0.00968905 | 546 | 9.39550974 | 58.11286615 | 546 | 18.5810923 | 29.3847095 |
| 7.2 | 27.3 | 0.00879162 | 459.25 | 6.93023404 | 66.26760329 | 458.75 | 14.0197726 | 32.7216435 |
| 8.4 | 29.1 | 0.00776222 | 337.5 | 4.12632647 | 81.79188009 | 337.25 | 9.48385609 | 35.5604299 |
| 9.6 | 84.9 | 0.00704467 | 259 | 2.54333709 | 101.8347121 | 258.75 | 6.3608808 | 40.6783287 |
| 9.9 | 33.3 | 0.00682288 | 243.5 | 2.28438056 | 106.5934478 | 243.5 | 5.80917686 | 41.9164377 |
| 10.5 | 39.6 | 0.00661816 | 216.75 | 1.87432642 | 115.6415436 | 216.5 | 5.01471726 | 43.1729226 |
| 11.4 | 49.5 | 0.0061692 | 184 | 1.38301841 | 133.0423358 | 184 | 4.00792491 | 45.9090437 |
| 12.6 | 63.9 | 0.0056585 | 150.75 | 0.94646636 | 159.2766589 | 150.75 | 3.01510339 | 49.9982855 |
| 15 | 8.4 | 0.00504859 | 106.75 | 0.49379416 | 216.1831981 | 106.75 | 1.94598341 | 54.8565829 |
| 18.6 | 22.8 | 0.00434646 | 70 | 0.23074235 | 303.3686681 | 69.75 | 1.1176789 | 62.406117 |
| 21.6 | 129.6 | 0.0039643 | 52 | 0.13437717 | 386.9704951 | 52 | 0.77937614 | 66.7200306 |
| 24.9 | 300 | 0.0036377 | 39.4 | 0.08633833 | 456.3442289 | 39.4 | 0.50946945 | 77.335354 |
| 27.3 | 78 | 0.00344266 | 32.75 | 0.06045672 | 541.709847 | 33 | 0.39805886 | 82.9023124 |

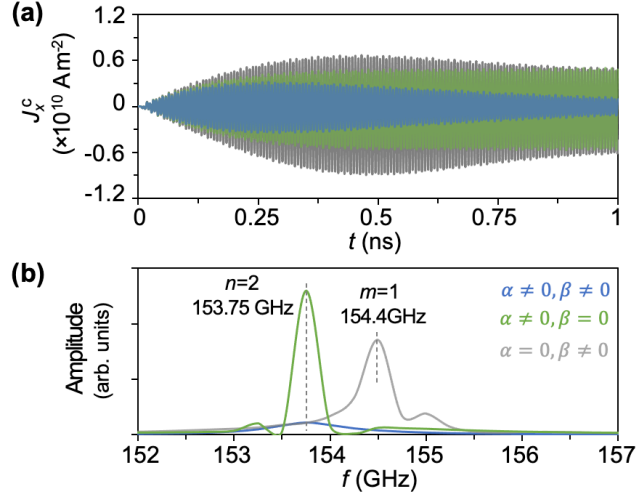

**Figure 14.** (a) Evolution of the total charge current $J_x^c(t)$ at the Pt/MAFO interface in a freestanding Pt(6.64 nm)/MAFO(12.45 nm)/SiN(31.54 nm) multilayer. $t=0$ ns is the moment the acoustic pulse was injected from the top surface of the Pt film. (b) Frequency spectra obtained by Fourier transform of the time-domain $J_x^c(t)$ data from $t$=0-4 ns. Blue curve: both magnetic and elastic damping are turned on ($\alpha \neq 0, \beta \neq 0$), with $Q$=154.6; gray curve: zero magnetic damping ($\alpha = 0, \beta \neq 0$), with $Q$=428.1; green curve: zero elastic damping ($\alpha \neq 0, \beta = 0$), with $Q$=588.0.

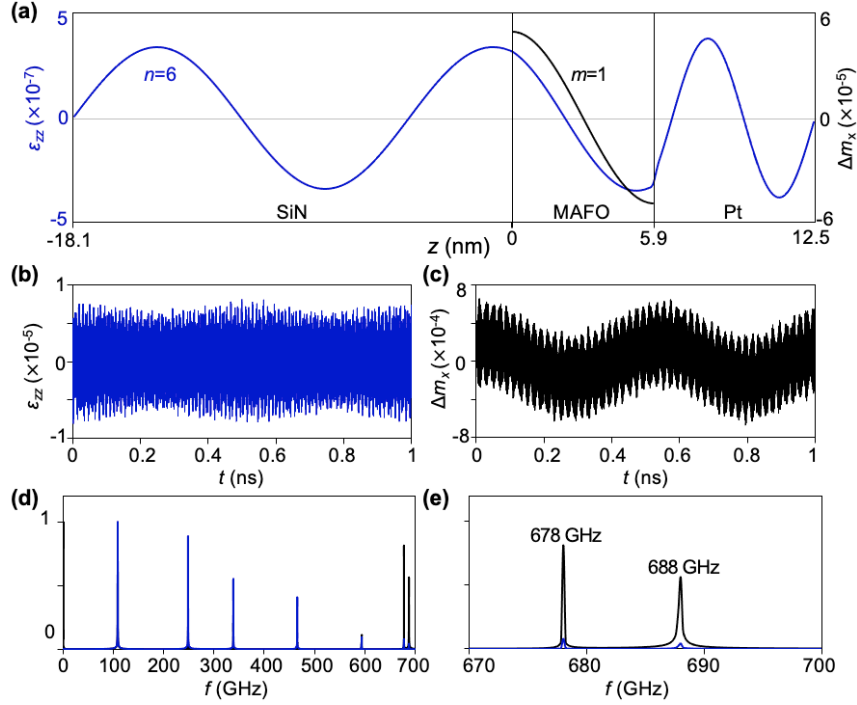

**Figure 15.** (a) Numerically extracted spatial profiles of the *n*=6 mode (blue) standing phonon mode $\varepsilon_{zz}$ in the Pt(6.6 nm)/MAFO(5.9 nm)/SiN(18.1nm) freestanding multilayer and the *m*=1 mode (black) magnon $\Delta m_x$ in the MAFO film. The spatial profiles are extracted by performing inverse Fourier transform of the 682.8 GHz peaks in their own frequency spectra with the backaction of the $\Delta \mathbf{m}$ on the $\varepsilon_{zz}$ being turned off. The $\varepsilon_{zz}$ and the $\Delta m_x$ have approximately the same spatial profile in the MAFO film. Evolution of (b) the strain $\varepsilon_{zz}(t)$, and (c) the magnetization change $\Delta m_x(t)$ at the Pt/MAFO interface. *t*=0 is the moment the acoustic pulse was injected from the top surface of the Pt film. Frequency spectra of both $\varepsilon_{zz}(t)$ and $\Delta m_x(t)$ in (d) 0-700 GHz, and (e) 670-700 GHz.